# Compact modeling technology for the simulation of integrated circuits based on graphene field-effect transistors

*Francisco Pasadas, Pedro C. Feijoo, Nikolaos Mavredakis, Aníbal Pacheco-Sanchez, Ferney A. Chaves, and David Jiménez\**

F. Pasadas, P. C. Feijoo, N. Mavredakis, A. Pacheco-Sanchez, F. A. Chaves, D. Jiménez*
Departament d'Enginyeria Electrònica, Escola d'Enginyeria, Universitat Autònoma de Barcelona, 08193 Bellaterra, Spain
E-mail: David.jimenez@uab.cat

F. Pasadas
Departamento de Electrónica y Tecnología de Computadores, Universidad de Granada, 18071 Granada, Spain.

In this study, we report the progress made towards the definition of a modular compact modeling technology for graphene field-effect transistors (GFET) that enables the electrical analysis of arbitrary GFET-based integrated circuits. A set of primary models embracing the main physical principles defines the ideal GFET response under DC, transient (time domain), AC (frequency domain), and noise (frequency domain) analysis. Other set of secondary models accounts for the GFET non-idealities, such as extrinsic-, short-channel-, trapping/detrapping-, self-heating-, and non-quasi static-effects, which could have a significant impact under static and/or dynamic operation. At both device and circuit levels, significant consistency is demonstrated between the simulation output and experimental data for relevant operating conditions. Additionally, we provide a perspective of the challenges during the scale up of the GFET modeling technology towards higher technology readiness levels while drawing a collaborative scenario among fabrication technology groups, modeling groups, and circuit designers.

## 1. Introduction

By combining graphene devices with interconnects and other components, innumerable circuits can be designed for analog and radio-frequency (RF) applications[1], preferably in the form of an integrated circuit (IC). The graphene-based circuits could be integrated with the silicon CMOS IC to increase the IC functionality. A possible realization of a hybrid graphene-silicon IC could consist of a silicon platform that would implement digital circuits according to the CMOS process. In addition to the silicon platform, a graphene platform (e.g., based on graphene transistor technology) for making RF functions exploiting the unique properties of graphene could be fabricated, combining the best of two technologies in a single IC (**Figure 1**a).[2] The graphene platform could be monolithically integrated on a convenient substrate (**Figure 1**b) such as silicon carbide[3] or a flexible polymer[4] to exploit the functionality of graphene in specific applications.[5] Regardless of the choice between hybrid or monolithic, a compact modeling technology for the graphene platform is required to make predictions of the electrical behavior of arbitrary circuits for DC, transient (time domain), AC (frequency domain), and noise (frequency domain) analysis. Every type of analysis requires a tailored set of models that capture the physics involved in a consistent way with experimental measurements for the relevant device operating conditions, which form the basis for a technology computer aided design (TCAD). Moreover, the mathematical complexity of the model should be kept low because various devices could be involved in a simulation. A TCAD tool is required during manufacturing to make the circuit design-fabrication cycle more efficient.

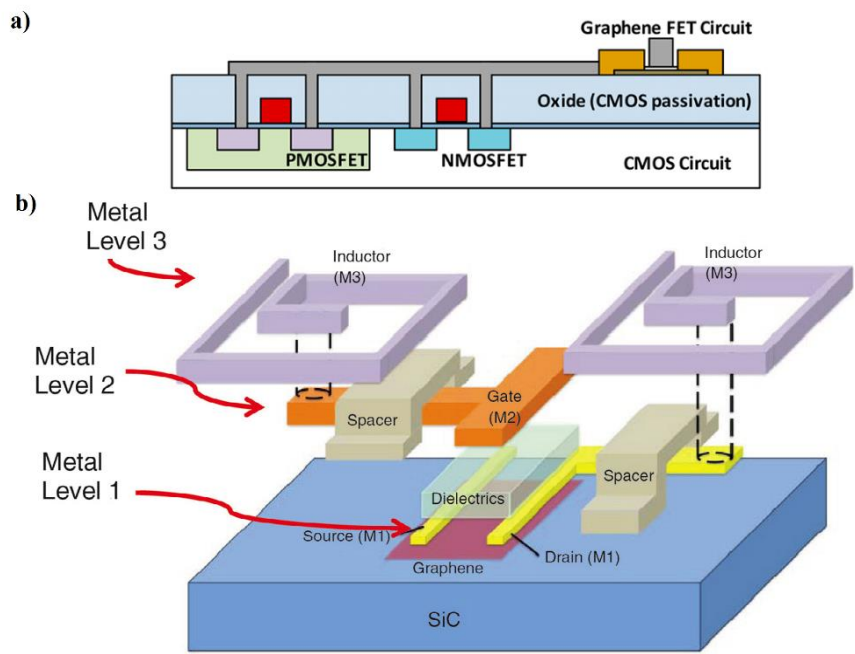

**Figure 1**. **a)** Schematic cross-section of a hybrid IC combining silicon and graphene

platforms, where the former implements the digital part of the mixed-signal circuit and the latter implements the RF part. Reproduced with permission.[2] 2016, ACS. **b)** A wafer-scale graphene circuit in which all circuit components, including GFET and inductors, are monolithically integrated on a single silicon carbide wafer. Reproduced with permission.[6] 2012, Elsevier.

In this study, we present the progress made towards the definition of a modular compact modeling technology for the graphene field-effect transistor (GFET), which is considered as the building block of the graphene platform in the IC. For a circuit simulation to be realistic, various non-idealities must be modeled at the device level. They include the effect caused by the parasitic network (parasitic resistances, capacitances, and inductances) associated to the contact pads and interconnections, as well as the effect of extrinsic resistances (contact and gate resistance), which significantly reduce the expected device performance. Additionally, the effect of the charges trapped/detrapped in the dielectric materials and interfaces must be included, which are related with the observed time-dependent drift of the operating bias point. In contrast, as short-channel length devices are needed to upscale the RF performance of graphene devices and circuits, short-channel effects (SCE) must be properly accounted by considering the two-dimensional (2D) electrostatics and velocity saturation effect. It is also important to consider the self-heating effects (SHE) at high fields that might produce an important reduction of the drain current. In addition, if the excitation frequency is close or above the device intrinsic cutoff frequency, the effect of carrier inertia should be considered, which requires a non-quasi static (NQS) model. For the noise, the relevant physics must be collected into an appropriate compact model to obtain the noise indicators at the circuit level. This requires an analysis of the different noise sources and their dependence on the substrate, dielectric environment, as well as contact technology. A deep understanding of the mechanisms of local noise propagation to terminal currents and voltages is necessary. At high-frequency (HF) range, a careful analysis of the coupling mechanisms between the channel and the gate(s) is required. These efforts are relevant because figures of merit are limited by noise.

A set of models capturing the relevant physics is needed to deal with the different types of electrical analysis, which forms the basis for a compact modeling technology targeting the simulation of ICs based on GFETs. As shown in **Figure 2**, they can be categorized into primary models defining the ideal device response and secondary models accounting for non-ideal effects to provide the necessary refinement to perform realistic circuit simulations. An in-depth explanation of relevant model details and main results for each type of analysis will be given along the Sections of the manuscript, namely DC (§2), transient (§3), AC (§4), and noise analysis (§5). In addition, we have benchmarked each model with experimental results to demonstrate its predictive capability.

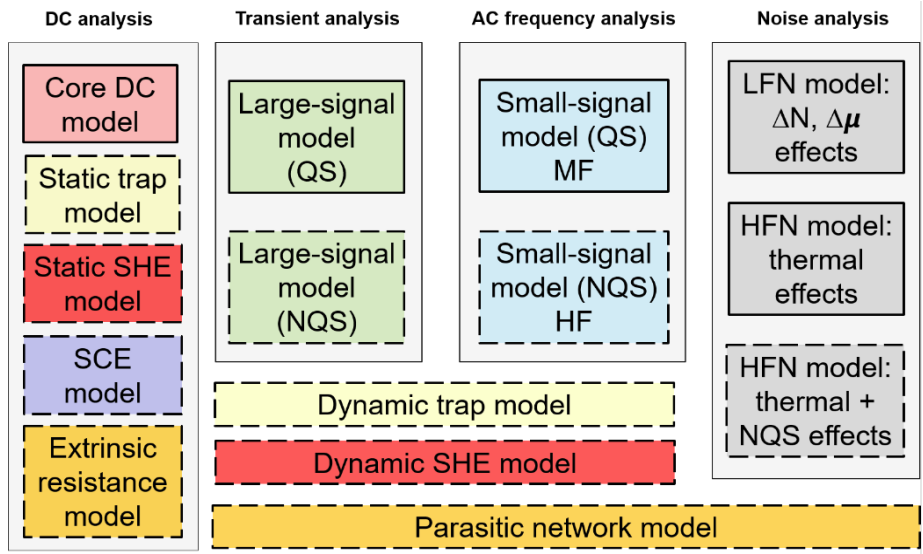

**Figure 2.** Modular compact modeling GFET technology presented in this study. The primary models describing the ideal device are represented by black solid lines, whereas the secondary models describing the device non-idealities are represented by black dashed lines. Acronyms used - SHE: self-heating effects, SCE: short-channel effects, QS: quasi-static, NQS: non-quasi-static, MF: medium frequency, HF: high frequency, LFN: low-frequency noise, HFN: high-frequency noise.

## 2. DC analysis

Here, a model accounting for the DC behavior of GFETs is presented. §2.1 introduces the intrinsic GFET core DC model, which is based on the Poisson's equation coupled with a drift-diffusion current equation, the former describing the device electrostatics (§2.1.1) and the latter describing the electron transport along the graphene channel (§2.1.2). In §2.2, various non-ideal effects affecting the operation of a GFET are added to the intrinsic model; thus, their inclusion is proven to be necessary to get consistency with experimental results from real devices. Those encompass charges trapped in the dielectric materials and interfaces (§2.2.1); SHE (§2.2.2); SCE (§2.2.3). Finally, some relevant analytical models for the contact resistance, which can be used towards extending the intrinsic model with the obtained values of this parameter, have been presented (§2.2.4).

2.1. Core DC Model

Here, we present the model for the electrostatics of a four-terminal GFET in §2.1.1, which sets the basis for the later formulation of a drain current model in §2.1.2.

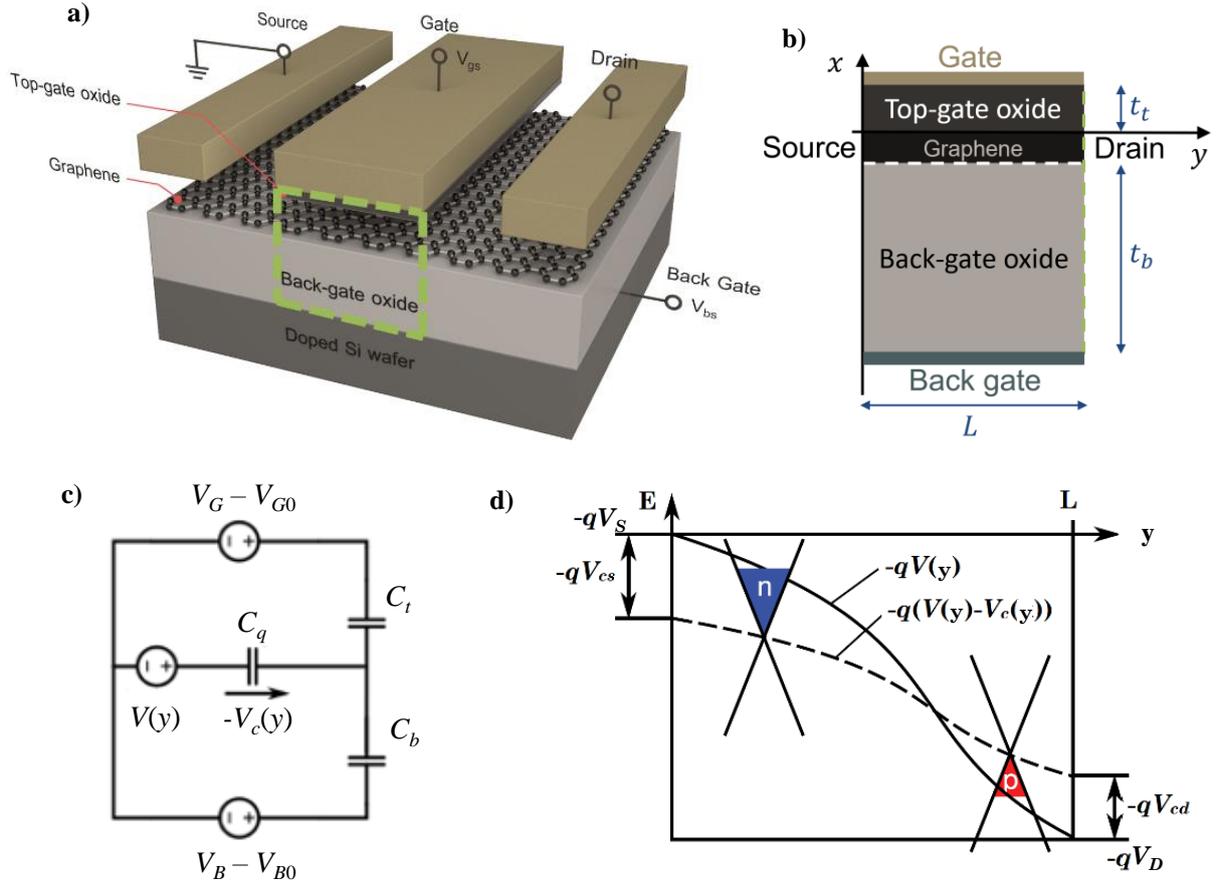

**Figure 3. a)** Schematic depiction of a four-terminal GFET. **b)** Cross-section of the GFET and the domain where the Poisson's equation is evaluated to solve the electrostatics. This area corresponds to the dashed rectangle in a). **c)** Equivalent capacitive circuit of the GFET. **d)** Schematic of the band diagram of the intrinsic device:[7,8] Energy, $E$, versus longitudinal position, $y$. The quasi-Fermi level $E_F = -qV(y)$ and the Dirac energy $E_D = -q\psi(y) = -q(V(y) - V_c(y))$ are shown. When the quasi-Fermi level is located at the Dirac energy, the dubbed Dirac point or charge neutrality point is reached. $V_D$ and $V_S$ are the drain and source biases, respectively; $V_{cd}$ and $V_{cs}$ are the channel potentials at the drain and source sides, respectively. Two Dirac cones illustrate the mixed $n/p$-type channel of this example. Reproduced with permission.[8] 2014, IEEE.[9] 2016, IOP.

### 2.1.1. Electrostatics of a Four-Terminal FET

Let us consider a general GFET as that depicted in **Figure 3**a. The graphene sheet under the electric control of the gate electrodes plays the role of the active channel. The electrostatic modulation of the carrier concentration in the 2D channel is achieved via a double-gate stack consisting of top- and back-gate dielectrics and corresponding metal gates (**Figure 3**b). The graphene sheet is contacted with both a drain and source terminal; carrier transport is produced between both terminals when a non-zero bias is applied. The direction of current transport defines the longitudinal direction ($y$). The transversal direction ($x$) goes from the top gate to the bottom gate, whereas the $z$-direction goes across the width of the device ($W$), where $W$ is sufficiently large such that the transistor can be considered uniform along the $z$ axis. The electrostatics of the GFET can be evaluated by applying the Poisson's equation across its structure

$$\nabla \cdot [\epsilon(x,y)\nabla\psi(x,y,T)] = \rho_{free}(x,y,T), \tag{1}$$

where $T$ is the temperature, $\epsilon$ is the permittivity, $\psi$ is the electrostatic potential, and $\rho_{free}$ is the free charge density. Considering a one-dimensional (1D) approximation for the GFET electrostatics along a transversal cut, the following charge-balance equation, which can be represented by the equivalent capacitive circuit shown in **Figure 3**c, can be obtained[10,11]:

$$\begin{aligned}Q_{net}(y,T) = &-C_t\big(V_G - V_{G0} - V(y,T) + V_c(y,T)\big) \\ &-C_b\big(V_B - V_{B0} - V(y,T) + V_c(y,T)\big),\end{aligned} \tag{2}$$

where $Q_{net}(y,T) = q[p(y,T) - n(y,T)]$ is the net sheet charge density; $q$ is the elementary charge; $p$ and $n$ are the hole and electron carrier densities, respectively. $C_t = \epsilon_t/t_t$ ($C_b = \epsilon_b/t_b$) is the top- (back-) oxide capacitance per unit area, with $\epsilon_t$ ($\epsilon_b$) the top (back) dielectric constant and $t_t$ ($t_b$) the top- (back-) oxide thickness. The top (back) overdrive voltage is $V_G - V_{G0}$ ($V_B - V_{B0}$), where $V_G$ ($V_B$) is the top- (back-) gate potential, and $V_{G0}$ ($V_{B0}$) comprises the work-function difference between the top (back) gate and the graphene channel as well as the effect of additional fixed charge owing to impurities or doping. Because of the presence of non-negligible contact resistances in GFETs (§2.2.4), the intrinsic source and drain potentials ($V_S$ and $V_D$ at the active channel edges) cannot be shorted to reference in practice; therefore, we adopted a general formulation where none of the terminals are grounded. Thus, the GFET is driven by terminal voltages defined with respect to some arbitrary reference point. $V(y,T)$ is the quasi-Fermi level along the graphene channel, i.e., the electrochemical potential, and it must fulfill the following boundary conditions: (i) $V(y = 0, T) = V_S$ and (ii) $V(y = L, T) = V_D$ at the source and drain edge-sides, respectively, where $L$ is the gate length. $V(y,T)$ is assumed to be the same for both electrons and holes because the generation/recombination times for carriers in graphene are very short (1100 ps-)[12–14]; therefore, electron and hole quasi-Fermi levels do not deviate significantly from each other.[7] $V_c(y,T)$ represents the position-dependent chemical potential, i.e. the shift of the Dirac potential ($\psi(y,T)$) with respect to the quasi-Fermi level. **Figure 3**d shows a scheme of the electrostatic, electrochemical, and chemical potentials along the channel at an arbitrary bias. Electron and hole concentrations can be calculated as a function of $V_c(y,T)$ using the following equations, which are a result of the density of states of graphene (deduced from its linear dispersion relation), where the electronic states are occupied according to the Fermi–Dirac statistics:

$$p(y,T) = \frac{\Delta^2}{2\pi(\hbar v_F)^2} + \frac{2}{\pi}\left(\frac{k_B T}{\hbar v_F}\right)^2 \mathfrak{F}_1\left[\frac{qV_c(y,T)}{k_B T}\right]$$

$$n(y,T) = \frac{\Delta^2}{2\pi(\hbar v_F)^2} + \frac{2}{\pi}\left(\frac{k_B T}{\hbar v_F}\right)^2 \mathfrak{F}_1\left[-\frac{qV_c(y,T)}{k_B T}\right], \quad (3)$$

where $\Delta$ is the amplitude of the electrostatic potential inhomogeneity that causes the electron-hole puddles;[15] $\mathfrak{F}_1$ is the Fermi–Dirac integral of first-order; $k_B$ is the Boltzmann constant, $\hbar$ is the reduced Planck constant, $v_F = \sqrt{3}a\gamma_0/2\hbar$ is the Fermi velocity; where $a = 2.49$ Å is the graphene lattice constant[16] and $\gamma_0 = 3.16$ eV is the interlayer coupling.[17] The term $\Delta^2/(2\pi(\hbar v_F)^2)$ accounts for the contribution of the puddles to the carrier concentration. The net sheet charge density and the quantum capacitance of graphene, which is defined as $C_q(y,T) = \partial Q_{net}/\partial V_c$, result in[8]

$$Q_{net}(y,T) = \frac{2q(k_B T)^2}{\pi(\hbar v_F)^2}\left(\mathfrak{F}_1\left[\frac{qV_c(y,T)}{k_B T}\right] - \mathfrak{F}_1\left[-\frac{qV_c(y,T)}{k_B T}\right]\right)$$

$$C_q(y,T) = \frac{2q^2 k_B T}{\pi(\hbar v_F)^2}\ln\left[2\left(1+\cosh\left[\frac{qV_c(y,T)}{k_B T}\right]\right)\right] \quad (4)$$

$V_c(y,T)$ could be calculated as a function of the terminal biases ($V_G$, $V_B$, $V_D$, $V_S$) from Equation 2 and 4. However, this formulation is not convenient for a compact model compatible with circuit simulators. Using a square-root-based approximation for $C_q$,[18] it is possible to get an implicit expression for $V_c$ that can be written in terms of elemental functions, which is more suitable for compact modeling purposes

$$C_q(y,T) \approx kc_1\sqrt{1+(V_c/c_1)^2}$$

$$Q_{net} = \int_0^{V_c} C_q(\tilde{V})\,d\tilde{V} = \frac{kc_1}{2}\left(V_c\sqrt{1+(V_c/c_1)^2} + c_1\,\mathrm{asinh}[V_c/c_1]\right), \quad (5)$$

where $k = 2q^3/(\pi(\hbar v_F)^2)$ and $c_1 = (k_B T/q)\ln(4)$.

The chemical potentials at the source $V_{cs} = V_c(0,T)|_{V=V_S}$ and drain $V_{cd} = V_c(L,T)|_{V=V_D}$ side-edges are the relevant quantities required to calculate the drain current. They can be easily determined by implementing the Verilog-A algorithm reported in,[8,19] which allows the circuit simulator to solve Equation 2 and 5, favoring a circuit-compatible model.

### 2.1.2. Drain Current Model

Here, we deal with the development of a compact model for the GFET drain current. For such purpose, we assumed that the drift-diffusion theory is applicable. That implies that the carrier mean free path (MFP) is significantly shorter than $L$. The MFP is related to the graphene quality and values below 100 nm have been reported at room temperature.[15] Electrons and holes in graphene tend to present mobilities in the same order of magnitude in both experiments and simulations[20–22]. Here, we assume an equal mobility for both types of carriers for simplicity, but the models could be easily extended for non-equal mobilities. The drain current of a GFET according to the drift-diffusion theory can then be evaluated as

$$I_{DS} = qW\rho_{sh}(y,T)\mu(y,T)\frac{\partial V(y,T)}{\partial y}, \tag{6}$$

where $\rho_{sh}(y,T) = p(y,T) + n(y,T)$ is the transport carrier sheet density and $\mu(y,T)$ is the field-dependent carrier mobility that reads as[9]

$$\mu(y,T) = \frac{\mu_{LF}}{\sqrt[\beta]{1+\left(\frac{\mu_{LF}}{v_{sat}(y,T)}\left|\frac{\partial \psi(0,y,T)}{\partial y}\right|\right)^{\beta}}}, \tag{7}$$

where $\beta$ is a parameter of the model describing how sharp the transition between low- and high-field mobilities is and $\mu_{LF}$ refers to the low-field carrier mobility. Saturation velocity $v_{sat}(y,T)$ is limited by optical phonon scattering according to the following equation:[23,24]

$$v_{sat}(y,T) = \frac{2\Omega}{\pi\sqrt{\pi\rho_{sh}(y,T)}}\frac{1}{N_{OP}(T)+1}\sqrt{1-\frac{\Omega^2}{4\pi v_F^2 \rho_{sh}(y,T)}}$$

$$N_{OP}(T) = \frac{1}{\exp\left[\frac{\hbar\Omega}{k_B T}\right]-1}, \tag{8}$$

where $N_{OP}(T)$ and $\hbar\Omega$ are the optical phonon occupation and energy, respectively. Low-field mobility and saturation velocity strongly depend on the dielectric materials surrounding the graphene layer, namely on the substrate and top-gate insulator used for the specific GFET geometry. These parameters depend on various scattering mechanisms that drive the carrier transport.

To obtain an explicit expression for Equation 6 in terms of the chemical potential, $V_c$, the following simplifications are needed: (i) under the condition of identical electron and hole mobilities, $\rho_{sh}(y,T)$ is approximated to its second-order Taylor expansion[8,18]

$$\rho_{sh}(y,T) \approx \frac{\Delta^2}{\pi(\hbar v_F)^2} + \frac{\pi(k_B T)^2}{3(\hbar v_F)^2} + \frac{q^2 V_c^2(y,T)}{\pi(\hbar v_F)^2} \tag{9}$$

(ii) A soft-saturation model ($\beta = 1$) is adopted. This value is consistent with numerical studies of electronic transport based on the Monte Carlo simulations;[25] (iii) $v_{sat}$ is assumed to be constant ($v_{sat,0}$), instead of using Equation 8. This is because the implementation of Equation 8 together with (ii) has been found to produce some artifacts that can result in harmonic distortion when large-signal transient simulations are performed. Considering all the simplifications, the following closed-form drain current equation is achieved:

$$I_{DS} = \frac{W\mu_{LF}}{L_{eff}}\left\{\frac{k}{2}\left(\frac{kc_1 V_c(c_1^2+2V_c+4c_2)\sqrt{1+(V_c/c_1)^2}}{8(C_t+C_b)} - \frac{kc_1^2(c_1^2-4c_2)\operatorname{asinh}[V_c/c_1]}{8(C_t+C_b)} + \frac{V_c^3}{3} + c_2 V_c\right)\right\}\Bigg|_{V_{cs}}^{V_{cd}}, \tag{10}$$

where $c_2 = (\pi k_B T)^2/3q^2 + \Delta^2/q^2$. An effective electrical length, $L_{eff}$, incorporating velocity saturation effects can be defined as

$$L_{eff} = L + \frac{1}{C_t + C_b}\frac{\mu_{LF}}{v_{sat,0}}\{Q_{net}\}_{V_{cs}}^{V_{cd}} \qquad (11)$$

2.2. Non-Ideal Effects for Enhancing the Prediction Capability of the Core DC Model

To achieve high-yield technology generations as well as reproducible electrical device characteristics towards exploiting GFETs at a circuit level, the trap mechanisms taking place within the device should be understood. Traps are material- or energetic-dependent imperfections within the channel or its surroundings, e.g., interfaces and oxide, able to capture and release carriers at different rates. Fast capture occurs within the channel and interfaces close to it, whereas the release of these trapped carriers is fast in the channel region and slow at the interfaces. In contrast, traps within the oxide region far from the channel have slow capture and emission time constants.[26] Graphene channels and their associated interfaces have been optimized towards a significant reduction of traps and defects in transistor architectures e.g., through passivation[27,28] and/or encapsulation techniques.[29,30] However, gate oxide has been a major issue in graphene transistors as well as in other emerging[26] and mature[31] FET technologies owing to scalability limitations;[26] thus, deep oxide (border) traps still affect the device performance.[32–35] High-κ insulators, such as $HfO_2$ and $Al_2O_3$, as well as layered 2D insulators, such as hexagonal boron nitride (hBN), have been used to fabricate high-performance GFETs.[36,37] Trap-related phenomena in the aforementioned bulk high-κ oxides have been observed and studied in silicon devices[38] as well as in emerging transistor technologies.[39] hBN has been suggested as the optimal dielectric for GFETs owing to the good lattice matching between the 2D channel and the 2D dielectric and minimized dangling bonds.[40–42] However, in addition to the low-κ of hBN (κ ≈ 3–4),[40,43] which limits the equivalent oxide thickness scalability,[26] causing wafer-scale integration issues,[36] few works have reported on HF GFETs with hBN as the true gate oxide,[29,44,45] which exhibit poor HF performance, in stark contrast to GFETs with high-κ oxides.[46–52] Furthermore, in contrast to the demonstrated low density of interface traps in hBN/graphene interfaces,[29,41] the temperature- and field-dependent carrier capture and emission processes owing to border traps within this 2D dielectric in 2D FETs,[26] specifically in GFETs, are not yet understood despite the recent characterization efforts.[53–55] A correct description of trap-related effects and their impact on the static and dynamic performance of graphene transistors can reveal true features of a corresponding technology.[34,35,56–58] The impact of traps on the transfer device characteristics is described by the static trap model module presented in §2.2.1, whereas the dynamic aspects of traps are introduced in §3.3.1 to complete the discussion.

The upscaling of RF performance in GFETs can be achieved by the progressive reduction of the channel length. When analyzing a short-channel GFET, velocity saturation is an important factor, but 2D electrostatics should also be considered. The strong electric fields caused at high bias affect the charge distribution along the graphene channel; thus, the 2D electrostatics across the plane perpendicular to graphene must be analyzed. Therefore, we developed a model that solves the 2D Poisson's equation coupled with the drift-diffusion equation, which includes the velocity saturation effect, to consider SCE effectively. In contrast, power dissipation in the graphene channel imperfectly spread out of the device increases its temperature,[59–61] triggering SHE. Temperature strongly affects charge transport through carrier concentration and saturation velocity. SHE and SCE will be modeled in §2.2.2 and §2.2.3, respectively.

Finally, in §2.2.4, we studied the bias-dependent contact resistance, which embraces the phenomena at the metal-graphene interface as well as the junction formed between the graphene layer under the metal electrode and the graphene channel.

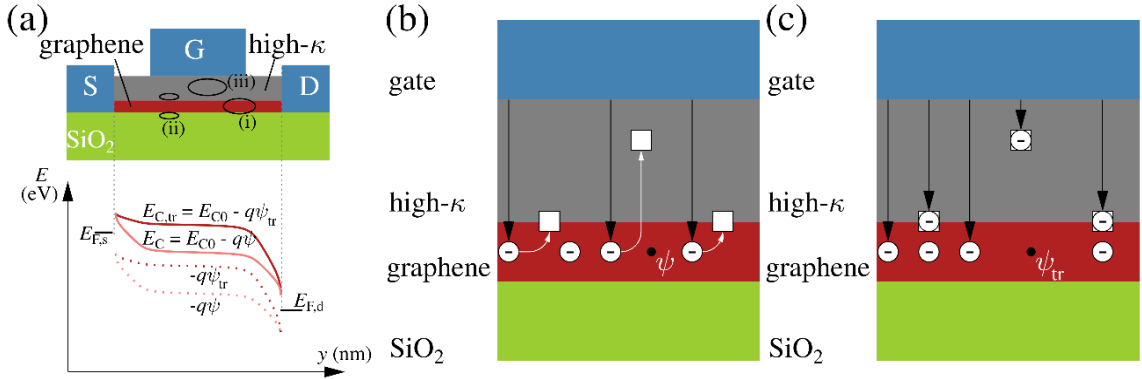

**Figure 4. a)** Top: schematic cross-section of a GFET indicating (i) channel, (ii) interface, and (iii) deep oxide traps. Bottom: sketch of the conduction energy band and channel potential under trap-free ($E_C$, $\psi$) and trap-affected ($E_{C,tr}$, $\psi_{tr}$) conditions. Schematic representation of the gated device region showing traps (squares) and carriers (circles) under **b)** trap-free and **c)** trap-affected conditions.

*2.2.1. Static Trap Model*

The actual performance of GFETs can be revealed by a device model correctly describing the hysteresis in both trap-affected and trap-reduced scenarios. In the literature, graphene transistor models including trap-effects have been reported;[34,58] however, their use has been limited either to observe the impact of the traps on the transport properties within a specific trap-affected scenario (forward bias sweep)[34] or without considering drain-to-source voltage dependence of the traps.[58] Next, we describe our model[35] that overcomes the latter issues.

The density of trap centers in a high-κ device, including GFETs, is higher in the gated region because all types of traps are present, i.e., material, interface, and border traps. Hence, the modeling approach reviewed next focuses on the specific device region. A high-κ oxide has been considered without loss of generality, i.e., trapping mechanisms in 2D oxides are qualitatively similar to the ones in 3D oxides but differ in temperature and vertical field dependence. Trapping mechanisms in ungated regions can be implicitly included in scattering-related parameters.

**Figure 4**a shows a schematic cross-section of a top-gated graphene transistor with high-κ oxide where trap centers are pointed out (top), and a sketch of the conduction band at a given $V_{GS}$ in the linear regime is also illustrated (bottom). At a hypothetical state of the traps not affecting the device performance, i.e., a trap-free state, the traps under the gate are empty; thus, the field lines emitted from the gate contact affect the channel carriers directly (cf. **Figure 4**b). Notably, a trap-free state is not possible in real conditions. However, a reduced impact of the charged traps on the *I-V* characteristics, i.e., minimum hysteresis window, can be obtained by controlling the measurement conditions with sophisticated characterization techniques, e.g., pulsed measurements.[32,35,56] In this study, the latter scenario is considered as a trap-reduced device performance. In a more practical scenario, the traps within the oxide are filled up so that the electrical field lines from the gate end on the traps, causing the channel potential to be shielded from the gate (cf. **Figure 4**c) and impacting the transport conditions, i.e., the operating bias point is shifted leading to a hysteretic device performance. Herein, the latter is named as a trap-affected performance.

To reproduce and understand the impact of traps on critical device figures of merit towards using the devices in circuits, a compact model describing the behavior of trap-affected and trap-reduced device performance accurately is required. The former characteristics can be obtained with a staircase characterization whereas an opposing-pulse technique can be used for the latter as shown in Ref.[35] Forward (increasing extrinsic, $V_{GS,e}$) and backward (decreasing $V_{GS,e}$) sweeps are applied in both characterization approaches to show a hysteresis window in the transfer characteristics. An analytical *I-V* compact model based on Equation 2 and 6,[10] under the condition of one active gate, has been used to describe the experimental data of GFETs by considering the impact of traps on the net charge within the graphene channel. The electrostatic equation of a practical one-gate GFET yields a net charge, $Q_{net,tr}(y,T)$, description with the impact of traps given as[35]

$$Q_{net,tr}(y,T) = -C_t[V_G - V_{G0} - V(y,T) + V_c(y,T)] + qN_{tr}, \tag{12}$$

where $N_{tr} = C_t(DV_{tr} - K_{tr}V_{DS}/2)/q$ is the trap density with the trap-induced potential term $DV_{tr} - K_{tr}V_{DS}/2$. $DV_{tr}$ and $K_{tr}$ are the phenomenological model parameters accounting for the shift of $V_{Dirac}$ owing to traps impact and the $V_{DS}$ dependence of this shift, respectively. The Dirac voltage, considering both the trap-reduced and trap-affected scenarios, is calculated by considering straightforward conditions such as $Q_{net,tr}(y,T) \to 0$ at the charge neutrality point (CNP) as well as an average channel potential over the channel, i.e., $V(y,T) \sim V_{DS}/2$ at similar bias conditions, which yields[35]

$$V_{Dirac} \approx \begin{cases} V_{G0} + \dfrac{V_{DS}}{2}, & \text{trap-reduced,} \\ V_{G0} - DV_{tr} + \dfrac{(K_{tr}+1)V_{DS}}{2}, & \text{trap-affected,} \end{cases} \tag{13}$$

To reproduce the traps impact on the performance, the net charge expression in the model should account the corresponding scenario, i.e., Equation 2 for the trap-reduced characteristics and Equation 12 for the trap-affected characteristics. Notably, the $V_{DS}$-dependence of a trap-affected Dirac voltage owing to hot-carriers[32,62] has been considered in this model,[35] which is an innovation with respect to other compact models including the impact of traps but neglecting such important effect.[58]

Both trap-affected and trap-reduced experimental transfer characteristics of a back-gated GFET technology with a gate width of 2 × 12 μm and a gate length of 300 nm (fabrication details in Ref.[63]), obtained via staircase sweep and an opposing sweep characterization, respectively,[35] have been correctly described by the model in the *p*-type region, as shown in **Figure 5**a-b (model parameters are listed in Table II of Ref.[35]). The hole dominant branch of the experimental data has been selected as the reference in the symmetric model, around $V_{Dirac}$, because better RF-related figures of merit are reported in this operation regime for this technology, in contrast to the *n*-type region.[35,64]

The model is valid for various $V_{DS}$ values, from low to higher ones, at different trap-affected states (**Figures 2** and **3** in Ref.[35]), as indicated by the correct $V_{DS}$-dependence of $V_{Dirac}$ observed in **Figure 5**c. The lower slope of the trap-reduced $V_{Dirac}$ versus $V_{DS}$ plot ($\sim 1/2$) with respect to the one obtained for the trap-affected data ($\sim (K_{tr}+1)/2$) suggests a trap-induced overestimation of $V_{Dirac}$ at high fields, in contrast to a reproducible performance with the traps impact highly suppressed, i.e., trap-reduced data. This is a crucial insight to be considered for the GFET performance in circuit applications.

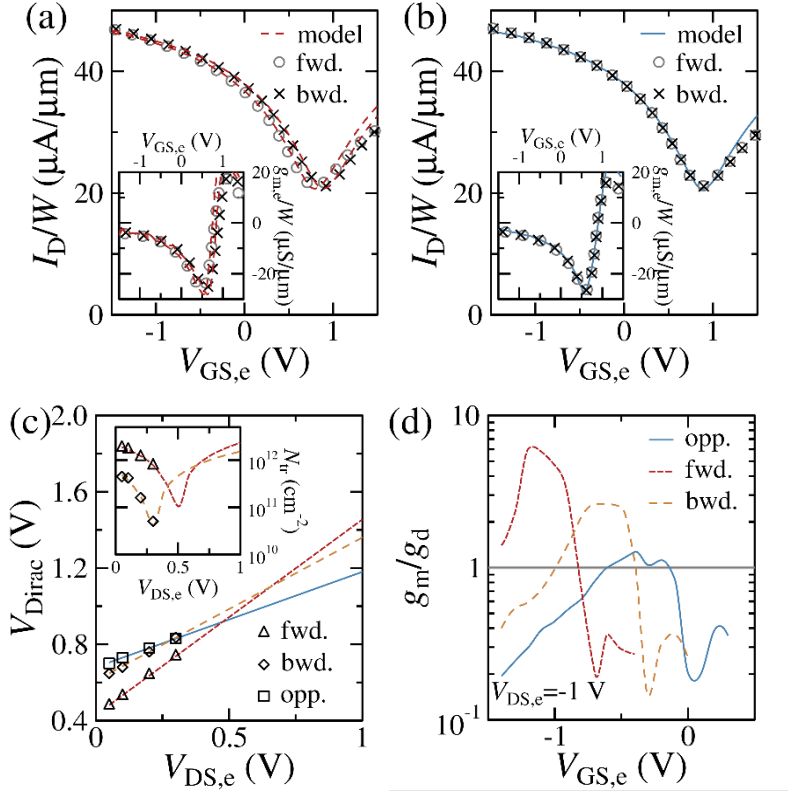

**Figure 5.** Experimental (markers) and modeling (lines) data of a 300 nm-long GFET: transfer characteristics with forward and backward sweeps showing **a)** trap-affected and **b)** trap-reduced behavior at $V_{DS} = 0.3$ V. Insets: transconductance plots at similar bias and conditions. **c)** $V_{DS}$-dependence of the Dirac voltage under different conditions. Inset: absolute value of the trap density for different $V_{DS}$. **d)** Intrinsic transistor gain at $V_{DS} = -1$ V. Adapted with permission.[35] 2020, IEEE.

Additionally, the trap density of the studied device is obtained for the different measurement sweeps from experimental data (see Ref.[35] for the procedure) and the model parameters. The model can describe the $V_{DS}$ dependence of $N_{tr}$, as shown in the inset of **Figure 5**c. Trapping and detrapping mechanisms can be elucidated from the curves for voltages lower and higher than the minimum point, respectively, which corresponds to a change of polarity of the trap-induced potential term. For instance, as indicated by the model curve, trapping processes are active for the forward staircase sweep at $0.3 \text{ V} < V_{DS} < 0.5$ V, whereas traps are enabled by the release of previous trapped carriers for the backward sweep case at the same bias range, as suggested by the increase of $N_{tr}$ for the model curve. A correct description of $N_{tr}$ at different biases, such as the one obtained with this model, enables the obtaining of improved insights in the device physics, e.g., on low-frequency noise (LFN) characterization of GFETs.[64]

The intrinsic gain, $A_{v,i} = g_m/g_{ds}$, calculated from both the intrinsic transconductance ($g_m = \partial I_{DS}/\partial V_G$) and output conductance ($g_{ds} = \partial I_{DS}/\partial V_D$), has been shown with the model calibrated with trap-affected and trap-reduced data (cf. **Figure 5**d) of the device under study.[35] The result indicates higher $A_{v,i}$ and different $V_{GS}$ dependence of this parameter for the trap-affected data than in the trap-reduced scenario. This is attributed to the trap-induced shielding of the channel potential because the changes of the current with respect to the different applied voltages, i.e., $g_m$ and $g_{ds}$, are affected differently depending on the traps state (measurement history). In addition to the experimental observations at different scenarios,[35] this model has shown the largely overlooked impact of traps on the performance of the graphene transistor. The approach used here enables a correct description of reproducible (trap-reduced) characteristics at room temperature towards exploiting such features in GFET-based circuit applications.

*2.2.2. Static Self-Heating Model*

Source-to-drain current within the graphene channel generates energy by the Joule effect, which can increase the temperature of the device considerably if the heat is not properly dissipated. The increase in graphene temperature, $T$, with respect to the ambient temperature, $T_A$, can be expressed as

$$T - T_A = \Re_{th} P_{dis}, \tag{14}$$

where $P_{dis} = |I_{DS} V_{DS}|$ corresponds to the power dissipated in the graphene channel and $\Re_{th}$ is the effective thermal resistance, which embraces all the paths through which the heat can be dissipated out of the channel. $\Re_{th}$ can be estimated according to the methodology proposed in Ref.[23,65]. The method considers the combined effect produced by the graphene/dielectric interface, dielectric layer, and substrate thermal conductance. The thermal conductance contributed by the contacts is neglected.

To demonstrate the impact of SHE, we simulated the device described in Ref.[51], assuming a high thermal resistance of 3·10⁴ K W⁻¹. **Figure 6**a shows the temperature distribution as a function of the bias. Temperatures can reach high values at large supply biases, in the order of the temperatures estimated for similar GFETs.[66] Moreover, experimental observations show that self-heating can increase the temperatures of graphene significantly when thermal resistances are large (for example, with thin insulators).[67] The transfer and output characteristics either neglecting SHE (solid lines) or considering SHE (dashed lines) are presented in **Figure 6**b-c. From the figure, it can be observed that temperature affects the drain current in different ways. The drain current increases in polarizations near $V_{Dirac}$, whereas it decreases for large gate voltages. Near $V_{Dirac}$, the increase in drain current is caused by a higher thermal carrier concentration. However, the temperature reduces carrier mobility and saturation velocity, whose effect dominates for biases far from $V_{Dirac}$. Moreover, current tends to saturate because of self-heating.[24] Despite the current saturation, §3.3.2 shows that the performance of the SHE-affected device observed in DC does not imply an improvement in the RF figures of merit, as the actual AC small-signal output conductance is larger than the one observed at DC.

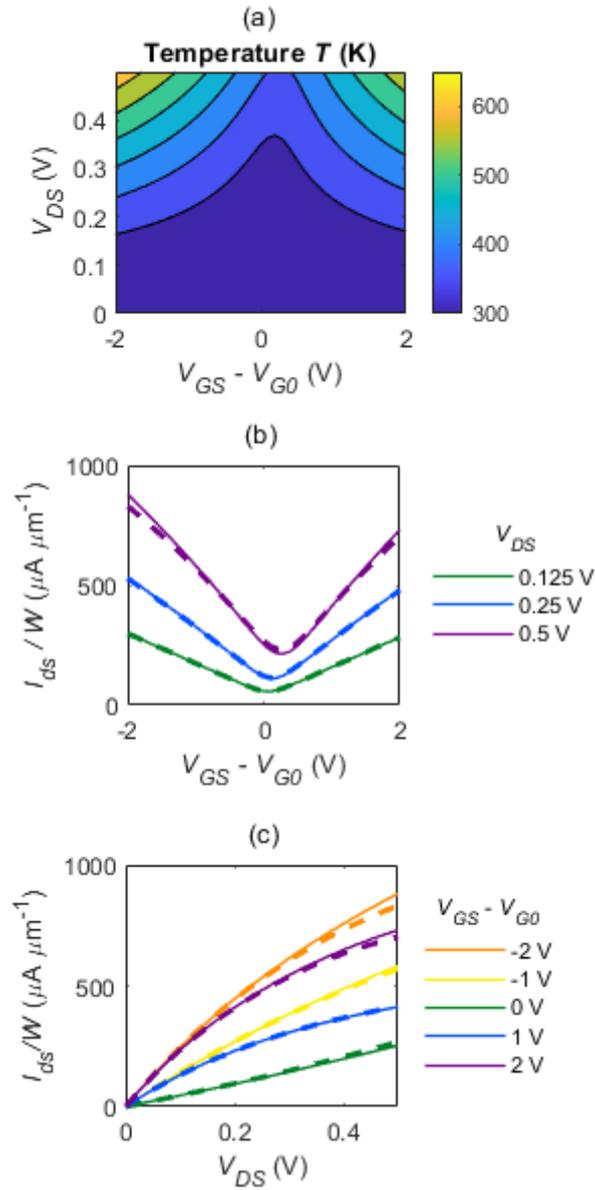

**Figure 6. a)** Temperature distribution as a function of the intrinsic bias voltage for a self-heated GFET. **b)** Transfer characteristics and **c)** output characteristics of a GFET at 300 K non-affected by SHE (solid) and affected by SHE (dashed).

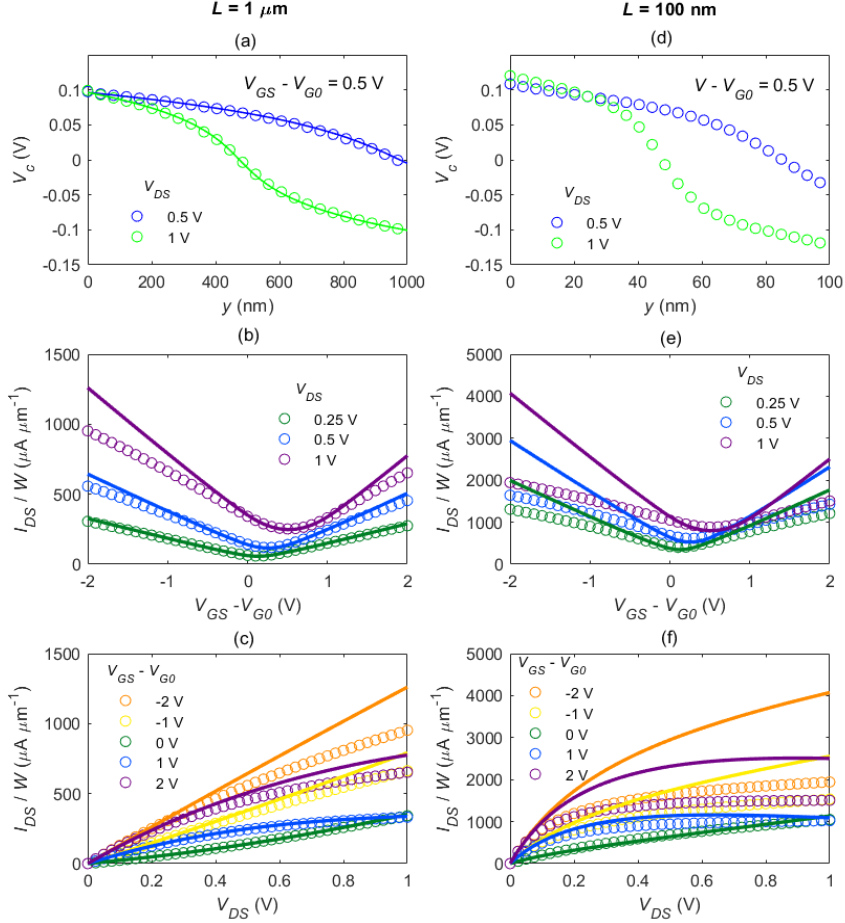

**Figure 7.** Comparison between the compact model (solid lines) and the numerical model that accounts for SCE (circles) for **a-c)** a 1 µm-long channel GFET and **d-f)** a 100 nm-long channel GFET. **a)** and **d)** Chemical potential along the channel. **b)** and **e)** Transfer characteristics. **c)** and **f)** Output characteristics.

*2.2.3. Short-Channel Model*

To study the performance of short-channel GFET, we developed a numerical model that aims to find the self-consistent solution of the 2D Poisson's equation (Equation 1) in the domain represented in **Figure 3**a and the current continuity equation (Equation 6). In addition to velocity saturation, this model accounts for the 2D electrostatics to consider SCE. Because Equation 6 assumes a drift-diffusion scheme, the short-channel model is valid when the GFET operates in this transport regime. Hence, the model cannot be applied to the pure ballistic regime.[68] This implies that the considered channel lengths should be significantly greater than the MFP of carriers in graphene, which is in the 10–100 nm range.[15,69,70]

We thoroughly studied SCE for the reference device reported in Ref.[51] by varying the channel length. **Figure 7** compares the compact model of §2.1 (solid lines) with the self-consistent model, which considers 2D electrostatics, (symbols). The figure shows the chemical potential and *I-V* curves for the cases of channel lengths of 1 μm, where both models give similar results, and 100 nm where models clearly diverge. The results of the short-channel device indicate that SCE imply a redistribution of carriers caused by 2D electrostatic effects:[9] carrier concentration close to source and drain edges increases as compared with the long device, as can be observed in the chemical potential distribution along the channel shown in **Figure 7**a and d. Additionally, the chemical potential slope becomes steeper, and the charge neutrality point (defined by the condition $V_c = 0$) is slightly displaced towards the middle of the channel.[9]

**Figure 7** also shows the differences between the compact model and the self-consistent model in predicting the transfer (**Figure 7**b and e) and output curves (**Figure 7**c and f). As for the 1 μm-long GFET, the differences between both models mainly come from the velocity saturation effect, where $v_{sat,0}$ is kept constant in the compact model whereas it follows the more complex model described by Equation 8 in the self-consistent model. However, the predicted trends are analogous. In contrast, the predictions of both methods differ considerably for the 100 nm-long GFET. For this case, a significant degradation in the transconductance can be observed, which has a direct impact on the RF performance scaling.

*2.2.4. Metal-Graphene Contact Resistance*

Although GFET has emerged as a promising device for analog/RF applications, the contact resistance ($R_c$) embracing the phenomena arising at the interface between graphene and source/drain metal electrodes remains a major limiting factor that affects the electronic transport properties.[71–80] For RF electronic applications, it is a relevant issue with a strong impact on figures of merit such as the maximum frequency of oscillation ($f_{max}$), intrinsic cut-off frequency ($f_T$), and $A_{v,i}$.[9,81,82] Despite the considerable number of experimental and theoretical studies, the origin of $R_c$ is still unclear owing to the multiple intrinsic and extrinsic factors affecting it, namely the nature of metals (chemisorbed or physisorbed), geometry of the contact (planar or edge), number of graphene layers, as well as impact of the fabrication process and bias. Therefore, a broad range of experimental values of $R_c$ have been reported in the literature even though the same contact metal has been considered.[83–93] In a metal-semiconductor junction (Schottky contact), a potential barrier (Schottky barrier) is formed at the interface. In an ideal case, the Schottky barrier height is given by the difference between the metal work function and the semiconductor electron affinity (Schottky–Mott approach). For conventional (3D) semiconductors, a Schottky contact can be turned into an ohmic contact by lowering the Schottky barrier height with opportune metal choice or lowering the barrier thickness sufficiently, in the order of few nanometers, by heavily doping the semiconductor. Therefore, electrons can go through the barrier driven by the quantum tunneling effect. However, for metal-graphene (MG) junction, the difference between the dimensional nature of the metal (3D) and graphene (2D) as well as the strong influence of the contact metal and the impact of the device fabrication details on the graphene properties hinder its description by the conventional Schottky–Mott picture. Furthermore, graphene zero energy bandgap prevents the formation of conventional Schottky contact and its small density of states near the Dirac point strongly limits the current injection from the metal.[94]

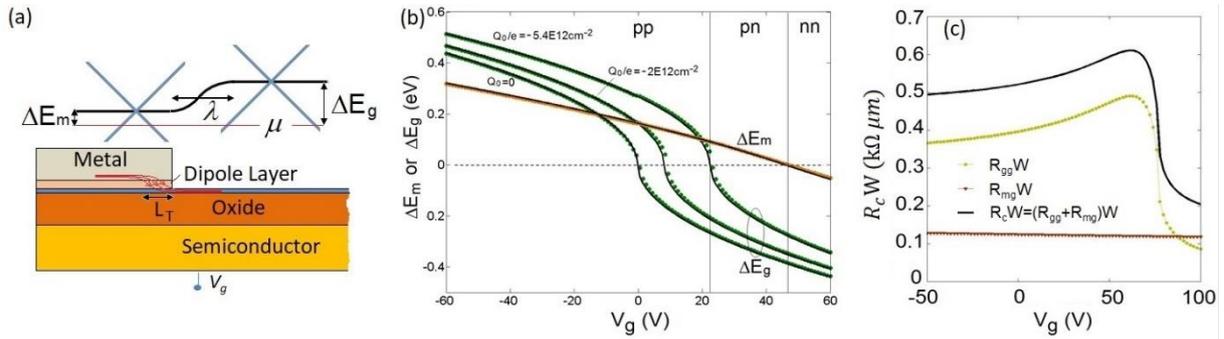

**Figure 8. a)** Sketch of the MG contact on a back-gate structure and the corresponding potential step formed between graphene under the metal and graphene in the channel. Red lines denote the current crowding effects near the contact edge. **b)** Graphene Fermi level shifts with respect to the Dirac point for different values of doping ($Q_0/q$), with palladium as a metal contact. **c)** Predicted $R_c$ and its components, $R_{mg}$ and $R_{gg}$, considering titanium as a metal. Reproduced with permission.[95] 2015, IOP.

When a finite metal electrode is deposited to cover part of a graphene sheet as in the GFET contact shown in **Figure 8**a, the following electrostatic effects appear: (i) there is a charge transfer through the interface producing a doping of the graphene underneath, which is characterized by a significant shift $\Delta E_m$ of the graphene Fermi level respect to its Dirac point. This is owing to the small density of states near the Dirac energy. Not only the difference between metal and graphene work functions should be considered in the process, but also the type of chemical interaction at the surface, namely physisorption or chemisorption.[96,97] (ii) A potential step is established between graphene under the metal and the graphene channel,[98,99] which is produced by a charge transfer between the two graphene regions. As shown in **Figure 8**a, the potential step is characterized by an effective length, $\lambda$, and energy shifts $\Delta E_m$ and $\Delta E_g$ in the graphene under the metal and graphene channel, respectively. In the limit where the contact length is larger than the transfer length, $L_T$, which is defined as the effective contact length contributing to the injection of carriers in graphene, there is a current crowding effect indicative of an $R_c$ dependence on the contact width instead of the contact area.[100,101]

Evidence of the current crowding has been reported by photocurrent spectroscopy experiment for graphene-gold contact.[102]

Apart from the above-mentioned electrostatics, several theoretical studies have been conducted to understand both intrinsic and extrinsic factors in controlling the values of $R_c$ in graphene-based devices. We can classify the theoretical studies as (i) ab-initio calculations[87,91,103–109] and (ii) analytical models.[78,95,96,101]. The ab-initio calculations have helped to comprehend the nature of the MG interface at the microscopic level and to suggest ways to engineer contact resistance. For example, they have been useful to explain the spread of $R_c$ measurements as due to uncontrolled graphene doping and/or the chemistry of the interface. However, these types of models have an extremely high computational cost. In contrast, for the analytical models, as proposed in Ref.[101], the transport in the MG junction is described as carrier injection from metal to graphene underneath with probability $\mathcal{T}_{mg}$ followed by injection to the graphene channel with probability $\mathcal{T}_{gg}$. The conductance of MG contact is expressed in terms of the conduction modes in graphene and the transmission probabilities following the Landauer approach. However, in this model, the mechanism of $\mathcal{T}_{mg}$ is ambiguous and it does not consider the 3D and 2D nature of the metal electrode and graphene sheet, respectively. Moreover, a theoretical model of carrier transport considering the dimensional nature has been developed in Ref.[95]. The physical model is based on the Bardeen transfer Hamiltonian (BTH) method for the calculation of the resistance, $R_{mg}$, between the metal and graphene underneath and on the Landauer approach to calculate the resistance, $R_{gg}$, arising at the potential step across the junction formed between the graphene under the metal and the graphene channel. The total contact resistance can be calculated as $R_c = R_{mg} + R_{gg}$. Here, the gate voltage dependence of $\Delta E_m$ and $\Delta E_g$, in a standard back-gate GFET configuration, is conveniently solved using a 1D model, as illustrated in **Figure 8**b, where palladium was assumed as the metal electrode. These two quantities are key factors to determine $R_c$ as a function of the gate voltage. Different types of junctions could be developed depending on the back-gate bias, namely *pp*-type, *pn*-type, and *nn*-type junctions. By assuming a chemical doping of the graphene channel $Q_0/q$ = -5.4 × $10^{12}$ cm$^{-2}$, the transitions from *pp*- to *pn*-type at $V_G$ ~ 23 V and from *pn*- to *nn*-type at $V_G$ ~ 46 V were captured in accordance with reported measurements.[101] Moreover, by combining the BTH method for calculating the specific contact resistivity,[96] $\rho_c$, with the transmission line model,[110] the simple analytical expression

$$R_{mg}(\Delta E_m) = \sqrt{\rho_c R_{sh}^m} \coth\left(\frac{L_C}{L_T}\right)/W_c \qquad (15)$$

is found; thus, its gate voltage dependence for arbitrary metals is predicted. Here, $R_{sh}^m$ is the sheet resistance of graphene under the metal; $L_c$ and $W_c$ are the contact dimensions. Additionally, $R_{gg}$ strictly depends on the effective length, $\lambda$, of the potential step that builds up between the graphene under the metal and the graphene channel, which can be calculated as

$$R_{gg}^{-1}(\Delta E_m, \Delta E_g) = 2e^2 W_c (h\pi)^{-1} \int_{-k_F}^{k_F} \mathcal{T}_{gg}\, dk_y, \qquad (16)$$

where $k_F = \min(|\Delta E_m|, |\Delta E_g|)/\hbar v_F$ and $\mathcal{T}_{gg}$ is the transmission probability of Dirac fermions across the potential step given by Cayssol et al.[98] According to this model, depending on the metal electrode and a possible chemical doping of the graphene channel, the two components of $R_c$ could be either similar in magnitude or of very different orders. It has been established that $R_{gg}$ is the dominant component for nickel and titanium electrodes, whereas there is a competition between both components for palladium. For illustrative purposes, the breakdown of $R_c$ in its two components for a titanium-contacted GFET is reproduced in **Figure 8**c.

Recently, a physical model of $R_c$ in titanium-contacted graphene-based FETs[78] has been developed considering an interfacial layer including oxidized Ti and polymethyl methacrylate (PMMA) residues at the Ti-G interface from the processing conditions of the contact. The study indicates that $R_c$ is highly dependent on the properties of the interfacial layer. Similar to Ref.[101], $R_c$ is calculated as

$$R_c^{-1} \propto \int_{-\infty}^{\infty} dE_1\, G(qE_1 - \Delta E_m, t_1) \int_{-\infty}^{\infty} dE_2\, G(qE_2 - \Delta E_g, t_2) \int_{-k_F}^{k_F} dk_y\, \mathcal{T}_{total} \qquad (17)$$

where

$$\mathcal{T}_{total} = \mathcal{T}_{mg}\mathcal{T}_{gg}/[1 - (1 - \mathcal{T}_{mg})(1 - \mathcal{T}_{gg})] \qquad (18)$$

is the total carrier transmission probability and the Gaussian functions $G(E, t)$ with effectives broadening $t_1$ and $t_2$ have been considered to get a more realistic model. Parameter $t_1$ considers the coupling between the metal and the quasi-bound graphene states underneath; $t_2$ is a parameter embedding information of the random disorder potential in the graphene channel, which depends on the minimum sheet carrier concentration.

The probability of carrier transmission through the potential is given in Ref.[98] The carrier transmission, $\mathcal{T}_{mg}$, through Ti-TiO$_x$-graphene interface depends on the tunneling mechanisms and it has been modeled based on quantum-mechanical tunneling theory using the WKB approximation.

The above comprehensive physics-based calculation of $R_c$ can be used as a confident reference for the values of this parameter in a device model. For compact modeling approaches, however, $R_c$ in GFETs has been either extracted from experimental data-based methodologies[75–77,79] (cf. §4.2.3) or been considered as a model parameter whose value is obtained by fitting the experimental I-V curves (cf. §3.2). Furthermore, parameter extraction techniques based on test-structures or analysis of transport experimental data[81,89,111,112] have been developed towards obtaining an immediate $R_c$ value. Further details of these extraction methodologies are provided in §4.2.3.

## 3. Transient Analysis

When a time-varying signal is applied to a GFET terminal, the dynamic operation of the device is strongly influenced by internal capacitive effects. Therefore, it is necessary to develop a large-signal model of intrinsic GFET capacitances. Various capacitive models for general FETs have been developed over the years. Basically, they can be categorized into two groups: (i) Meyer-like[113] and (ii) charge-based capacitance models. The advantages and shortcomings of both model groups applied to conventional FETs have been widely discussed and implemented in circuit simulators.[114,115]

Meyer-like models are widely used in incumbent technologies because of their simplicity and fast computation. They assume that the intrinsic capacitances are reciprocal (i.e., $C_{ij} = C_{ji}$). Notably, this hypothesis might give unphysical results when dealing with some class of circuits (such as switched capacitor filters); furthermore, earlier models based on this assumption could not ensure charge conservation.[116,117]

Conversely, the charge-based models ensure both charge conservation and nonreciprocity of intrinsic capacitances in a FET. Owing to some corrections assembled by Ward and Dutton,[118] the charge-conservation issue was solved by introducing a capacitive matrix, which adds a bit of complexity. In Ref.[11,119–121], we have provided graphs of bias-dependent $C_{dg}$ and $C_{gd}$ for graphene- and MoS$_2$-based FETs, showing that the reciprocity of capacitances in emergent 2D technologies cannot be assumed for all transistor operation regimes. However, most of the GFET capacitance models reported so far rely on the Meyer's reciprocity assumption without evaluating the implications of adopting it.[122–126] Therefore, a comparison of the RF performance prediction between the Meyer-like models against a charge-based approach is provided in Ref.[119], indicating that significant errors could arise.

Considering all the aforementioned, we presented a large-signal model for the GFET elsewhere and summarized it in §3.1.[11,127] Thereafter, to illustrate the working of the model, we have presented the dynamic response of several circuits exploiting the graphene ambipolarity in §3.2. Later in §3.3, we considered significant non-ideal effects that must be modeled to enhance the predictive capability of the large-signal model. The effects include the effect of the trapped charges in §3.3.1 and the NQS effects owing to carrier inertia in §3.3.2. Dynamic SHE will be treated in the context of small-signal analysis later in §4.2.2. As for the large-signal case, dynamic SHE would need some adaptation, which is not yet investigated. Finally, to account for the impact of parasitic elements when a transient circuit analysis is performed, a parasitic network model of the GFET should be considered in the simulation. That will be discussed in the context of AC analysis in §4.2.3.

3.1. Large-Signal Model

Assuming a quasi-static (QS) operation, the entering terminal currents in the time domain can be expressed by considering that the charges per unit area at any time controlled by the time-varying terminal voltages, ($v_G(t)$, $v_B(t)$, $v_D(t)$, and $v_S(t)$), are identical to those found if DC voltages ($V_G$, $V_B$, $V_D$, and $V_S$) were used.

$$i_m(t) = \frac{dQ_m}{dt} + I_m, \qquad (19)$$

where $m$ stands for $G, D, S$, and $B$, and they denote the top gate, drain, source, and back gate, respectively. The possible leakage current through the top and back insulators is neglected, that is, $I_G = I_B = 0$. In addition, to guarantee charge conservation, $I_D = -I_S = I_{DS}$, where $I_{DS}$ is calculated according to Equation 10. A four-terminal FET can be modeled using a set of 16 intrinsic capacitances, including 4 self-capacitances and 12 transcapacitances. The capacitance matrix is formed by these capacitances, where each element, $C_{ij}$, describes the dependence of the charge at terminal $i$ with respect to a varying voltage applied to terminal $j$, assuming that the voltage at any other terminal remains constant. Therefore, Equation 19 can be written as

$$\begin{bmatrix} i_G \\ i_D \\ i_S \\ i_B \end{bmatrix} = \begin{bmatrix} C_{gg} & -C_{gd} & -C_{gs} & -C_{gb} \\ -C_{dg} & C_{dd} & -C_{ds} & -C_{db} \\ -C_{sg} & -C_{sd} & C_{ss} & -C_{sb} \\ -C_{bg} & -C_{bd} & -C_{bs} & C_{bb} \end{bmatrix} \begin{bmatrix} dv_G/dt \\ dv_D/dt \\ dv_S/dt \\ dv_B/dt \end{bmatrix} + \begin{bmatrix} 0 \\ I_{DS} \\ -I_{DS} \\ 0 \end{bmatrix}, \qquad (20)$$

$$C_{ij} = -\frac{\partial Q_i}{\partial V_j}, \ i \neq j; \quad C_{ij} = \frac{\partial Q_i}{\partial V_j}, \ i = j. \quad i,j = G, D, S, B$$

Each row must sum to zero for the matrix to be reference-independent, and each column must sum to zero for the device description to be charge-conservative.[128] Notably, only 9 out of the 16 intrinsic capacitances are independent. In addition, we can take advantage of the relations between top- and back-gate capacitances,[11,129] namely $C_{bd} = C_{gd}(C_b/C_t)$; $C_{bs} = C_{gs}(C_b/C_t)$; $C_{db} = C_{dg}(C_b/C_t)$; $C_{sb} = C_{sg}(C_b/C_t)$; $C_{gg} = -C_{bg}(C_t/C_b) + C_tWL$; $C_{bg} = C_{gb} = -C_{bb}(C_t/C_b) + C_tWL$; reducing the independent set of intrinsic capacitances of a four-terminal GFET to only four; for instance, $C_{gs}$, $C_{gd}$, $C_{dg}$, and $C_{sd}$.

Considering all the aforementioned, the modeling of the dynamic response of a GFET requires a charge model relating the terminal charges to the terminal voltages. From the electrostatics given in Equation 2, the following relations are derived:[11]

$$Q_G + Q_B = -W \int_0^L Q_{net}(y)dy; \begin{cases} Q_G = Q_0 - \frac{C_tW}{C_t+C_b} \int_0^L Q_{net}(y)dy \\ -Q_B = Q_0 + \frac{C_bW}{C_t+C_b} \int_0^L Q_{net}(y)dy \end{cases}, \quad (21)$$

where $Q_0 = WLC_tC_b(V_G - V_{G0} - V_B + V_{B0})/(C_t + C_b)$. The charge controlled by the drain and source terminals is calculated based on the Ward–Dutton linear charge partition scheme, which guarantees charge conservation[118]

$$Q_D = W \int_0^L \frac{y}{L} Q_{net}(y)dy$$

$$Q_S = -(Q_G + Q_B + Q_D) = W \int_0^L \left(1 - \frac{y}{L}\right) Q_{net}(y)dy \quad (22)$$

Both terminal charges and intrinsic capacitances can conveniently be written using $V_c$ as the integration variable, as it was done to model $I_{DS}$ (cf. §2.1.2). Because $I_{DS}$ is the same at any point $y$ in the channel (by considering compensated generation/recombination processes), the equations required to evaluate the terminal charges and intrinsic capacitances are obtained from the drift-diffusion transport model as follows:

$$dy = \frac{\mu_{LF}W}{I_{DS}} \rho_{sh}(V_c)\frac{dV}{dV_c}dV_c - \frac{\mu_{LF}}{v_{sat}}\frac{d\psi}{dV_c}|dV_c|$$

$$y = \frac{\mu_{LF}W}{I_{DS}}\left(\int_{V_{cs}}^{V_c} \rho_{sh}(V_c)\frac{dV}{dV_c}dV_c\right) - \frac{\mu_{LF}}{v_{sat}}\left|\int_{V_{cs}}^{V_c} \frac{d\psi}{dV_c}dV_c\right|, \quad (23)$$

where $\frac{dV}{dV_c} = 1 + \frac{d\psi}{dV_c} = 1 + \frac{C_q(V_c)}{C_t+C_b}$.

The drain current model presented in §2.1.2 is combined with the charge-based compact intrinsic capacitance description as shown in Equation 20 to assemble a large-signal model of GFETs. The modeling approach presented here has been validated against numerical simulations (**Figure 9**c)[11] as well as with experimental data from fabricated GFETs.[127]

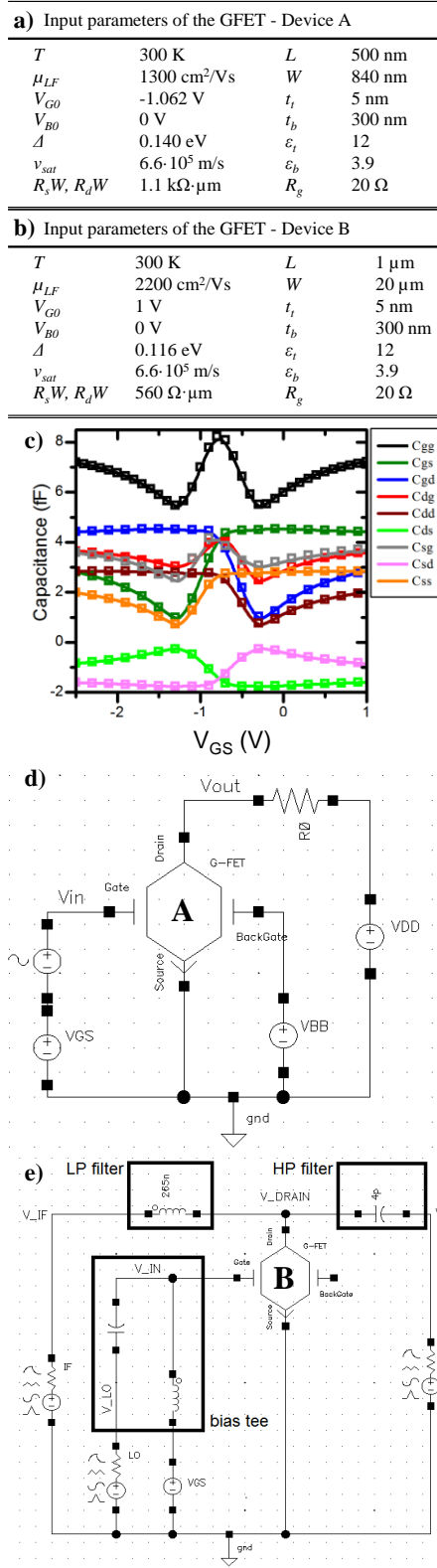

**Figure 9.** Input parameters used to describe the GFETs reported in **a)**[130] (Device A) and **b)**[131] (Device B). Contact resistances, $R_d$ and $R_s$, have been considered bias-independent. $R_g$ is the gate resistance. **c)** Compact model (solid lines) and numerical (symbols) calculation of the intrinsic capacitances versus $V_{GS}$ for Device A at $V_{DS} = 1$ V. **d)** Schematic circuit of a GFET-based frequency doubler based on Device A.[130] **e)** Schematic circuit of a subharmonic resistive GFET mixer based on Device B.[131] A bias tee is used for setting the DC bias point. The characteristic impedance is 50 Ω. Reproduced with permission.[11,127] 2016, IEEE.

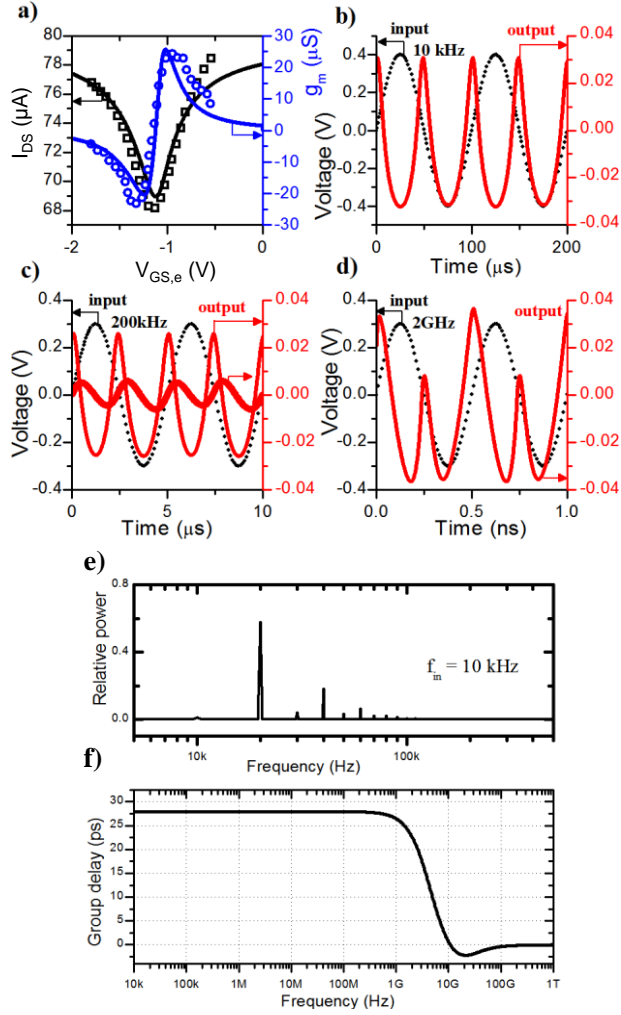

**Figure 10. a)** Experimental measurements (symbols) and simulations (solid lines) of the DC transfer characteristics and the extrinsic transconductance of a GFET-based frequency doubler.[130] The device is biased at $V_{DS,e} = 1$ V, $V_{BS,e} = 40$ V, and $V_{GS,e} = -1.15$ V. **b)** Input and output waveforms considering an input frequency of $f_{in} = 10$ kHz and amplitude $A = 400$ mV. **c)** Input and output waveforms considering an input frequency of $f_{in} = 200$ kHz and amplitude of $A = 300$ mV. A thicker solid line shows the output waveform when a parasitic capacitance ($C_{pad} = 600$ pF) is placed between the drain/source and the back gate, considering the effect of the electrode pads. **d)** Input and output waveforms considering an input frequency of $f_{in} = 2$ GHz and amplitude of $A = 300$ mV. **e)** Power spectrum obtained via Fourier transforming the output signal shown in b). **f)** Group delay vs. frequency of the GFET-based frequency doubler. Reproduced with permission.[127] 2016, IEEE.

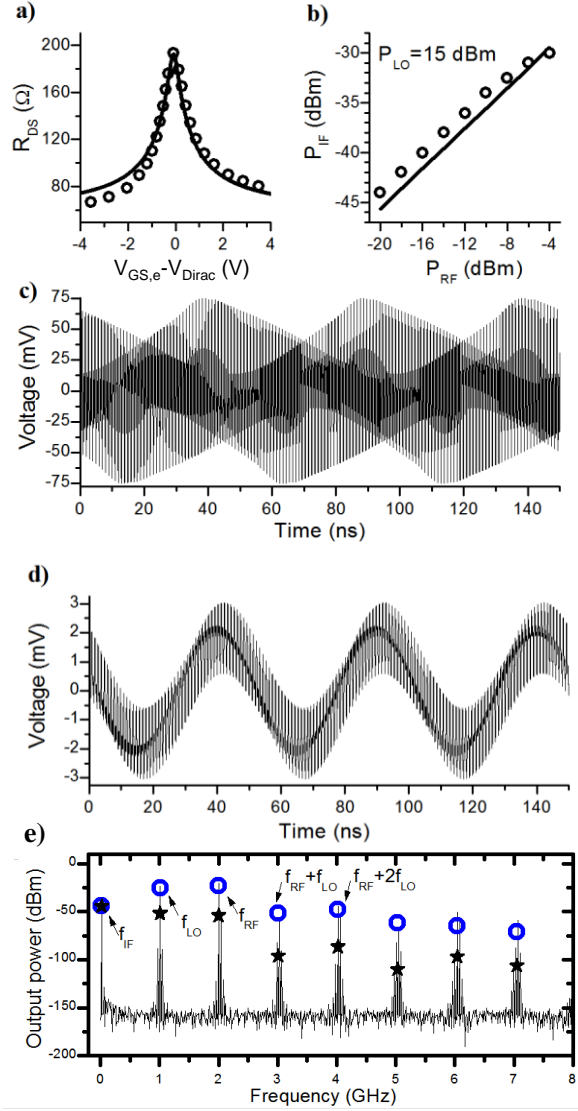

**Figure 11. a)** Drain-to-source resistance, $R_{DS}$, versus the overdrive gate voltage, with $R_{DS} = R_d + R_s + R_{ch}$, where $R_{ch}$ is the channel resistance; $R_d$ and $R_s$ are the extrinsic contact resistances at the drain and source sides, respectively. Solid lines correspond to simulations and the symbols to the experimental results.[131] **b)** IF output power as a function of the RF input power. The device is biased at the Dirac voltage $V_{GS,e} = V_{Dirac}$ and $P_{LO} = 15$ dBm. **c)** Transient evolution of the signal collected at the drain at $V_{GS,e} = V_{Dirac}$. The following conditions have been assumed: $P_{LO} = 15$ dBm and $f_{LO} = 1.01$ GHz; $P_{RF} = -20$ dBm and $f_{RF} = 2$ GHz. **d)** Transient evolution of the IF signal collected at the IF port under the same conditions as in c). The separation between peaks is 50 ns, which corresponds to $f_{IF} = |f_{RF} - 2f_{LO}| = 20$ MHz. **e)** Simulated spectrum (solid lines) of the signal collected at the drain (shown in c) and the measured power peaks (blue circles) reported in Ref.[131] The stars correspond to the simulation results of power peaks of the signal collected at the IF port (shown in d). Reproduced with permission.[127] 2016, IEEE.

## 3.2. Dynamic Response of GFET-based RF Applications

To demonstrate the predictive capabilities of the large-signal model presented,[127] we consider various exemplary circuits, namely: the frequency doubler and the subharmonic mixer.[130,131] The GFET used in each of the circuits is referred as device A and B (with their corresponding parameters given in **Figure 9**a-b), respectively. In **Figure 9**c, a set of independent intrinsic capacitances for the device A is plotted as a function of the gate voltage, showing a clear non-linear behavior of the GFET around the Dirac voltage. It can be observed that the Meyer's reciprocity does not hold.[11]

A graphene-based frequency doubler leverages the quadratic-like transfer characteristic ($I_{DS}$ vs. $V_{GS,e}$ curve) of a GFET. If such a transfer characteristic is not perfectly parabolic and/or symmetric, which is the practical case, the output voltage contains the doubled frequency and other higher-order harmonics, resulting in harmonic distortion. As for device A configured in the topology depicted in **Figure 9**d, with the nearly symmetric transfer characteristic around $V_{Dirac} = -1.15$ V shown in **Figure 10**a, the output waveforms after considering different input frequencies are shown in **Figure 10**b-d. For the input signal with amplitude *A* and lowest frequency, $f_{in} = 10$ kHz, the output waveform consists of the doubled frequency with amplitude $\sim A/10$, with a clear distortion coming from other higher order harmonics (**Figure 10**b). A Fourier transform of such waveform is shown in **Figure 10**e, revealing that 60% of the output RF power is concentrated at the doubled frequency of 20 kHz.

When the input signal is increased up to $f_{in} = 200$ kHz and beyond, a significant decay of the output signal amplitude was observed in the experiment,[130] with a voltage gain of $\sim A/100$ likely because of the presence of a parasitic capacitance (estimated in $C_{pad} = 600$ pF) between the GFET source-drain terminals and its back gate. When including the $C_{pad}$, the predicted output waveform is similar to that in the experiment for an input frequency of 200 kHz (**Figure 10**c). If the input frequency is increased further to 2 GHz, as shown in **Figure 10**d, the output waveform displays the doubled frequency, although with a greater distortion because the group delay is not constant with the frequency according to **Figure 10**f, indicating that the phase is not linear with the frequency.

Furthermore, with the use of a single GFET (**Figure 9**e), a subharmonic mixer can be designed to take advantage of the device nonlinearity. This way, the device is fed with two different frequencies (the local oscillator, LO, signal at $f_{LO}$ and the RF signal at $f_{RF}$) and a mixture of several frequencies appears at the output port, including both original input frequencies; the sum of the input frequencies; the difference between the input frequencies; the intermediate frequency (IF), $f_{IF}$; and other intermodulations.[132] According to the graphene-based mixer topology shown in **Figure 9**e, the LO signal and DC bias are applied to the gate port through a bias-tee, whereas the RF signal is applied to the drain of the GFET through a high-pass filter, and the IF is extracted with a low-pass filter, both assumed with cut-off frequencies of 800 and 30 MHz, respectively.

The drain-to-source resistance, $R_{DS} = V_{DS,e}/I_{DS}$, versus the gate bias is shown in **Figure 11**a. To reach subharmonic operation, the device is biased at $V_{GS,e} = V_{Dirac} = 1$ V through a bias tee. In contrast, **Figure 11**b shows the mixer IF output power versus the RF input power, where a near constant conversion loss rate of ~25 dB is obtained. The transient evolution of the signal collected at the drain is shown in **Figure 11**c, as well as the signal collected at the IF port (**Figure 11**d), which oscillates as expected at $f_{IF} = |f_{RF} - 2f_{LO}| = 20$ MHz given that $f_{LO} = 1.01$ GHz and $f_{RF} = 2$ GHz. Finally, the spectrum of the signal collected at the drain is presented in **Figure 11**e, with an output power of $\sim -49$ dBm. Lower levels of odd harmonics are observed as well, which are attributed to the non-perfect symmetry of $R_{DS}$ versus $V_{GS,e}$.

3.3. Non-Ideal Effects for Enhancing the Large-Signal Model Prediction Capability

A brief description of the dynamic trap-related phenomenon in GFETs is presented in §3.3.1. The implementation of this module into the GFET modeling framework presented in this work is still an ongoing effort; hence, the corresponding subsection should be considered a useful guide for approaching this development. A systematic study on NQS effects in GFETs is provided in §3.3.2.

*3.3.1. Dynamic Trap Model*

Trapping and detrapping dynamic processes in MOS-like transistors can be characterized by their corresponding capture and emission time constants, $\tau_c$ and $\tau_e$, respectively. These phenomena can occur over a wide time span depending on the location of the traps centers within the device (cf. §2.2.1), as well as on the bias and temperature conditions.[133,134] In devices with high-κ oxides, e.g., GFETs, the dynamic performance is strongly affected by the capture and release of carriers,[35,135,136] mainly located at the gate oxide and channel interfaces, because the trap-induced shift of the channel potential modify the bias conditions (transfer curve hysteresis) to achieve specific dynamic characteristics.[35,136] Additionally, trapping phenomena have an impact on the device measurement history, i.e., on the initial state of shielding of the channel potential from the gate-source voltage. This impact could not be reduced by a technology-dependent quiescent time approach,[51,81] if either $\tau_c$ or $\tau_e$ (or a combination of both) is larger than the quiescent pulse duration. Hence, the dynamic modeling approach for graphene transistors should consider the trap time constants for a correct description and projection of the device performance. Furthermore, a reliable compact model considering a description for trap mechanisms can aid the technology development by revealing the device dynamic performance affected by traps under different pulse biasing conditions, as demonstrated in other emerging technologies.[137,138] The latter can be immediately exploited in specific application scenarios, e.g., high-speed GFET-based modulators designs,[139,140] and it can be boosted by an accurate modeling of the dependence of the trap-affected/reduced device performance on the pulse biasing conditions.

Trap dynamics in graphene transistors have been systematically characterized[141,142] using silicon technology-based models[143] yielding the traps activation energy distributions and trap time constant distributions of the capture and emission processes. Individual values of trap-related time constants have been experimentally characterized in different GFET technologies ranging from few nanoseconds to the order of few hours.[32,33,144–146] Trap time constants have been obtained by fitting the transient current response over specific non-quiescent conditions with constant pulse widths using empirical exponential models.[32,33,144,145] A capture process is usually related to such extracted trap time constants in these cases; hence, the emission time constants are ignored. Studies with different pulse duration reveal both $\tau_c$ and $\tau_e$, as demonstrated in Ref.[146], where the latter is obtained with pulses larger than the time required for an apparent initial steady-state of the current.

Currently, physics-based models for $\tau_c$ and $\tau_e$ in GFETs have not been reported in the literature. However, previously developed high-κ dielectric MOS-models of oxide traps, e.g., a non-radiative multi-phonon model[147] (already applied to other 2D transistor technologies),[148] can be adapted to GFETs because statistical similarities between trapping processes in the latter and in incumbent technologies have already been observed.[141,142] This model predicts $\tau_c$ and $\tau_e$ as a function of the position and energy of the trap, applied bias, and temperature.

A numerical device simulation solution, enabled by the drift-diffusion-based description of transport in graphene transistors,[149,150] can be developed by considering the trap-assisted phenomena, including $\tau_{c/e}$–dependent capture and emission rates, in both the Poisson's equation and the continuity equation[151,152] as already implemented in other drift-diffusion-based simulators describing different transistor technologies.[152–155] In contrast, an immediate approach to adapt the compact graphene transistor model considering traps in a static regime described in §2.2.1 for the dynamic description can be conducted by defining the trap density $N_{tr}$ in Equation 12 in terms of the steady-state trap density $N_{tr,ss}$ along with $\tau_c$ or $\tau_e$, as implemented previously in a different compact model.[156] The challenging characterization of $N_{tr,ss}$ has been overcome in Ref.[156] using an empirical function depending on the vertical fields and some fitting parameters. An improvement of the compact GFET model discussed in this work (cf. §2.1), including the dependence of both lateral and vertical fields as well as $\tau_c$ and $\tau_e$ in the definition of $N_{tr}$ is left for future studies. Alternatively, for circuit design purposes, a practical approach considering an adjunct trap network, as demonstrated in studies of emerging transistor technologies,[58,137,138,157] with different time constants defined by $RC$ networks can be an option to include directly in the compact model the impact of the different capture and emission trapping processes within the device.[39,137,138]

*3.3.2. Non-Quasi-Static Large-Signal Model*

Depending on the input frequency of the time-varying signal, two operating regimes can be distinguished, QS and NQS. In the QS regime, the fluctuation of the varying terminal voltages is sufficiently slow such that the channel charge can follow the voltage variations. This regime applies whenever the transition time for the voltage to change is larger than the transit time of the carriers from source to drain. Contrarily, the NQS regime where carrier inertia effects are important should be considered. When dealing with circuit simulations, assuming a QS regime is not appropriate for long-channel GFETs operating at HFs or when the load capacitance is extremely small.[128,158] Applications of QS approach could result in important errors when predicting phase margins or the stability of wideband amplifiers.[159]

A straightforward approach for modeling a transistor at speeds where the QS regime breaks down involves splitting the channel length in many shorter sections; thus, the QS approach still holds within each section.[128,160,161] To track the breakdown of the QS regime, we consider the frequency-dependent admittance $y_m = y_{dg} - y_{gd}$ as a convenient indicator of the electrical gate control on the transistor channel over frequency. Thus, a decrease of $|y_m|$ is interpreted as the loss of the gate control over the channel charge because of the significant carrier inertia originated at HF.[128,162] Considering the 1 $\mu$m-long prototype GFET described in **Figure 12**a, a simulation of normalized $|y_m|$ from the QS model is shown in **Figure 13**f. The selected bias to perform the calculation is $V_G = 1.5$ V, $V_D = 1$ V and $V_B = V_S = 0$ V. The result is compared with that obtained after connecting 20 identical GFETs of length $L/20 = 50$ nm in series (sharing all the same gate, as shown in the inset of **Figure 13**f), where the array effectively allows the capture of NQS effects. At medium frequencies (MF), both approaches, QS and NQS, predict the same normalized $|y_m|$. However, the upward-going magnitude of normalized $|y_m|$ predicted by the QS model for the 1 $\mu$m-long GFET working at HF is clearly unrealistic, as it suggests an enhancement in the forward gate-to-drain action. This behavior results are contrary to the expectation that, at such frequencies, control of the gate on the drain current is gradually lost due to the carrier inertia in the graphene channel.

These predictions for GFETs are in qualitative agreement with the NQS studies that have been conducted for conventional silicon-based MOSFETs.[128,159]

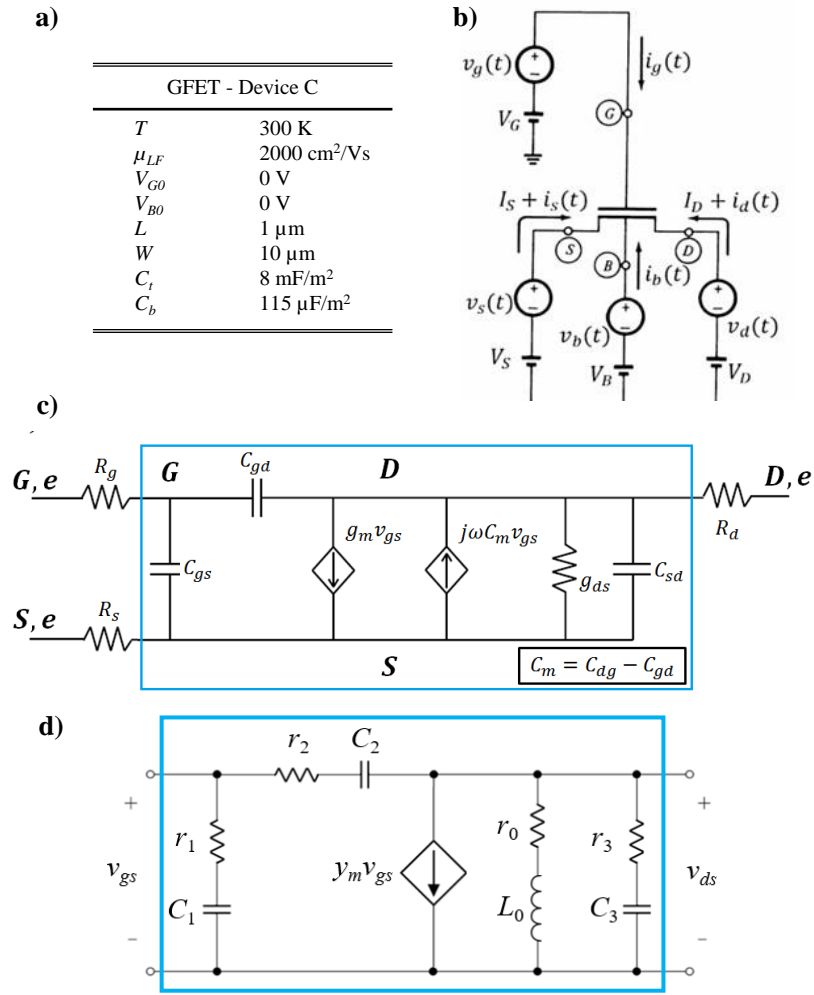

**Figure 12. a)** Input parameters used to describe a prototype GFET. **b)** Schematic of a four-terminal FET operating under small-signal regime showing the terminal DC and AC voltages as well as currents. To guarantee charge conservation, the sum of the terminal currents must be zero; thus, $I_D = -I_S$, where the DC top- and back-gate currents are $I_G = I_B = 0$; and $i_g(t) + i_b(t) + i_d(t) + i_s(t) = 0$. **c)** QS charge-based small-signal model suited to three-terminal GFETs. The equivalent circuit of the intrinsic device is framed in blue. **d)** Equivalent circuit of a GFET in two-port configuration describing the first-order NQS behavior using lumped elements. Reproduced with permission.[119,162] 2017 and 2020, IEEE.

## 4. AC Analysis

When considering analog and RF electronic applications, FET terminals are polarized with a DC bias over which a time-varying voltage is superimposed (**Figure 12**b). If the voltage amplitude is sufficiently small, the resulting AC components of terminal currents and charges can be linearly related to the AC voltage.[128] Therefore, the nonlinear FET can be treated as a linear circuit formed by lumped elements, known as the FET small-signal equivalent circuit. In §4.1, we derive a small-signal equivalent circuit for the GFET assuming the QS hypothesis valid in the MF range. Later in §4.2, we introduce some important non-ideal effects. Specifically, the small-signal model derivation has been extended to cover the NQS regime at HF in §4.2.1. A brief review of the dynamic self-heating model in GFETs is presented in §4.2.2. Finally, in §4.2.3, we discuss a procedure adapted to GFETs to go from the extrinsic to the intrinsic AC frequency response, which requires a model for the parasitic network that should include the effect of the contact pads, metal interconnections, and substrate.

4.1. Quasi-Static Small-Signal Model

Currently, many small-signal models proposed for GFETs have been directly imported from Meyer-like capacitance models.[123,124,126,163,164] They assume that the intrinsic capacitances are reciprocal, which has been proven to result in inaccuracies when predicting the RF figures of merit, as demonstrated in Ref.[119] for GFETs. Moreover, these models do not ensure charge conservation, which is crucial not only for accurate device modeling and circuit simulation, but also for proper parameter extraction. To ensure both capacitance reciprocity and charge conservation, the charge-based small-signal model shown in **Figure 12**c is adopted, where the intrinsic part has been framed in blue. We considered a three-terminal GFET configuration, where the back gate is considered AC disconnected ($i_B(t) = 0$). The small-signal parameters include $g_m$; $g_{ds}$; and a set of independent capacitances, including $C_{gd}$, $C_{gs}$, $C_{sd}$, and $C_{dg}$, are calculated according to the procedure described in §3.1. The intrinsic model must be augmented with the extrinsic resistances, $R_d$ and $R_s$, after linearization around the operating bias point, and the gate resistance, $R_g$ (**Figure 12**c), if the extrinsic RF performance is intended.

The small-signal model parameters could be either extracted by applying a direct methodology based on S-parameter measurements[119] or numerically computed with a DC simulator (optionally including SCE and/or SHE), allowing the prediction of RF performance for different embodiments of GFETs at arbitrary bias.[24,120,150]

4.2. Non-Ideal Effects for Enhancing the Prediction Capability of the Small-Signal Model

Here, we present a set of secondary models to enhance the prediction capability of the small-signal model presented in §4.1 by including some relevant non-idealities. We first describe a NQS small-signal model suitable for HF simulations in §4.2.1. Thereafter, a dynamic SHE model, developed in the context of small-signal analysis, is presented in §4.2.2. Finally, a parasitic network model that considers the effect of contact pads, metal interconnections, and the substrate is discussed in §4.2.3.

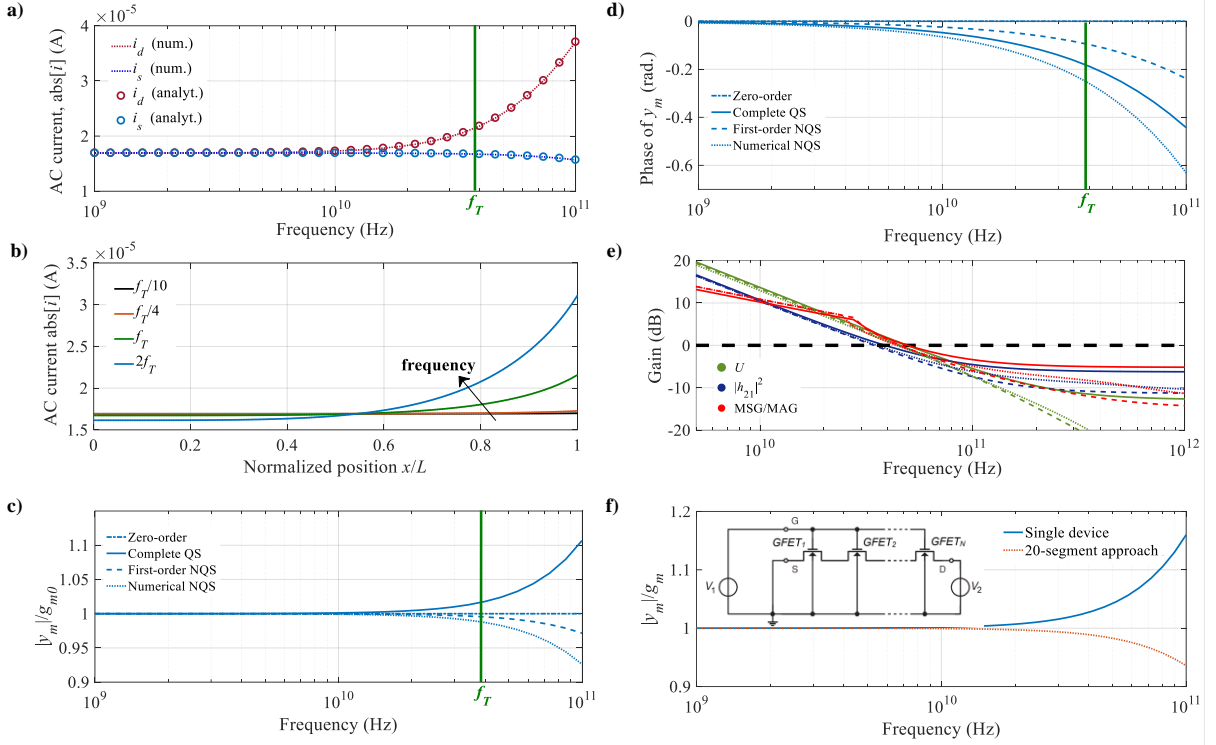

**Figure 13. a)** Modulus of the AC current at the drain ($i_d$) and source ($i_s$) edges computed in analytical (symbols) and numerical (dotted line) ways. The former is computed by truncating the modified Bessel functions of the first kind to $n = 10$ (Ref.[162]). **b)** Modulus of the AC current along the normalized position in the channel for different frequencies. **c)** Normalized magnitude and **d)** phase of $y_m = y_{dg} - y_{gd}$ versus frequency at $V_G = 1.5$ V, $V_D = 1$ V, $V_B = V_S = 0$ V. Four types of models are considered: zero-order model (solid black lines); charge-based QS model described in §4.1; first-order NQS model described in §4.2 (dashed lines); numerical NQS model (dotted lines). **e)** Small-signal current gain ($h_{21}$), unilateral power gain ($U$, Mason's invariant)[165] and maximum stable gain / maximum available gain (MSG/MAG) versus frequency of the GFET under test predicted by the QS (solid lines), first-order NQS (dashed lines), and numerical NQS (dotted lines) models. The RF figures of merit $f_T$ and $f_{max}$ are obtained when the gains are reduced to unity (0 dB). **f)** Normalized magnitude of $y_m$ versus frequency under the operating bias point $V_G = 1.5$ V, $V_D = 1$ V, $V_B = V_S = 0$ V for a single 1 µm-length GFET compared against a two-port configuration of a cascade of 20 GFETs, 50 nm-length each, connected in series. (inset) Schematics of the multi-segment approach applied to a GFET. Reproduced with permission.[162] 2020, IEEE.

*4.2.1. Non-Quasi-Static Small-Signal Model*

Here, we discuss ways for expanding the frequency range that can be reached with the small-signal model while keeping the accuracy to acceptable levels. Therefore, NQS effects arising at HF should be incorporated into the model. A small-signal NQS model can be derived from the analytical solution of the drift-diffusion equation coupled with the continuity equation, which can be expressed in terms of the modified Bessel functions of the first kind.[162] The model can be conveniently simplified to provide an equivalent circuit of lumped elements that can be used in circuit simulators. For instance, if second- and higher-order terms in $\omega$ (angular frequency) are neglected in the GFET $\omega$-dependent admittances, a first-order NQS equivalent circuit of the GFET can be obtained (**Figure 12**d), which is fully described by the following bias-dependent small-signal parameters:[162]

$$C_1 = g_{m0}(\gamma \tau_4 - \tau_2); \quad C_2 = g_{m0}\tau_2; \quad C_3 = g_{ds0}\tau_3(1-\gamma)$$
$$r_1 = \tau_1/C_1; \quad r_2 = \tau_1/C_2; \quad r_3 = \tau_1/C_3 \qquad (24)$$
$$L_0 = \tau_1/g_{ds0}; \quad r_0 = 1/g_{ds0}; \quad y_m = g_{m0}/(1+j\omega\tau_1)$$

where:

$$g_{m0} = -\mu_{LF}\frac{W}{L}\frac{k^2}{2(C_t+C_b)}\left(\text{Sign}[V_{cd0}]h_{GD}V_{cd0}{}^3 - \text{Sign}[V_{cs0}]h_{GS}V_{cs0}{}^3\right)$$

$$g_{ds0} = \mu_{LF}\frac{W}{L}\frac{k^2}{2(C_t+C_b)}(h_{GD}+h_{BD})|V_{cd0}|^3 \qquad (25)$$

$$V_{cx0} = \frac{(C_t+C_b)-\sqrt{(C_t+C_b)^2 \pm 2k[C_t(V_G-V_{G0}-V_X)+C_b(V_B-V_{B0}-V_X)]}}{\pm k}$$

$$h_{GX} = -\frac{C_t}{(-C_t-C_b \pm kV_{cx0})}; \quad h_{BX} = \frac{C_b}{C_t}h_{GX},$$

where the positive (negative) sign applies when $C_t[V_G - V_{G0} - V_X] + C_b[V_B - V_{B0} - V_X] < 0 (> 0)$ and the subscript $X$ stands for drain ($D$) and source ($S$). The time constants $\tau_1, \tau_2, \tau_3,$ and $\tau_4$ from Equation 24 are

$$\tau_1 = -\frac{D'}{12}\frac{V_{cd0}{}^4 V_{cs0}{}^4}{V_{cd0}{}^2 + V_{cs0}{}^2};$$

$$\tau_2 = \frac{D'}{24}\frac{(V_{cd0}{}^6 - 3V_{cd0}{}^2 V_{cs0}{}^4 + 2V_{cs0}{}^6)}{h_{GD}V_{cd0}{}^3 - h_{GS}V_{cs0}{}^3}h_{GD}V_{cd0}{}^3;$$

$$\tau_3 = -\frac{D'}{24}\left(V_{cd0}{}^6 - 3V_{cd0}{}^2 V_{cs0}{}^4 + 2V_{cs0}{}^6\right);$$

$$\tau_4 = \frac{D'}{24}\frac{(V_{cd0}{}^2 - V_{cs0}{}^2)^2}{h_{GD}V_{cd0}{}^3 - h_{GS}V_{cs0}{}^3}\left(h_{GD}V_{cd0}{}^3(V_{cd0}{}^2 + 2V_{cs0}{}^2)\right. \qquad (26)$$
$$\left. + h_{GS}V_{cs0}{}^3(2V_{cd0}{}^2 + 2V_{cs0}{}^2)\right)$$

$$D' = \text{Sign}[V_{cd0}]\mu_{LF}W^2k^3/\left(2I_{DS0}{}^2(C_t+C_b)\right)$$

$$I_{DS0} = \mu_{LF}Wk^2\left(\text{Sign}[V_{cd0}]V_{cd0}{}^4 - \text{Sign}[V_{cs0}]V_{cs0}{}^4\right)/(8L(C_t+C_b))$$

with $\gamma = C_t/(C_t+C_b)$.

Because of the simplifications made in the model derivation, the selected bias point should be placed far enough from the Dirac voltage for the device to operate in the linear region. The assumption is $V_{cx0}{}^2 \gg \left(\pi k_B T/(\sqrt{3}q)\right)^2 + 2q\Delta^2/(k\pi(\hbar v_F)^2)$. Although it is certainly a limitation of the model, the approach is useful for the important usage of GFET as an amplifier. In addition, the expressions shown in Equation 25 and 26 consider $k|V_{cx0}| \gg (C_t + C_b)$. The opposite case, $(k|V_{cx,0}| \ll (C_t + C_b))$, is provided in Ref.[162].

To demonstrate the impact of NQS effects on the RF performance of GFETs, we considered the device described in **Figure 12**a. It consists of a double-gated topology with 10 nm-Al$_2$O$_3$ and 300 nm-SiO$_2$ dielectrics at the top and back gate stack, respectively. **Figure 13**a shows the absolute value of the small-signal current at the drain ($i_d$) and source ($i_s$) edges for several frequencies, as well as the calculated $f_T$ (= 38.8 GHz). Both $i_d$ and $i_s$ have been computed both in an analytical and numerical way through the evaluation of the modified Bessel functions of the first kind.[162] The analytical case consists of truncating the function to a tenth order. According to **Figure 13**a, $|i_d|$ and $|i_s|$ show same value for frequencies lower than $\sim f_T/4$, which agrees with the QS assumption and with the behavior of silicon based FETs operating at MF.[128] For frequencies higher than $f_T/4$, $|i_d|$ and $|i_s|$ differ from the QS value adopting different values, which indicates that the channel charge in the graphene layer cannot follow the voltage variations for such frequencies.

**Figure 13**b shows the modulus of the AC current, $|i|$, along the normalized channel length for different frequencies. The solid orange line shown in **Figure 13**b corresponds to the frequency of $f_T/4$, which has been chosen to delimit the QS and NQS regimes.[128] For frequencies lower than $f_T/4$, e.g., $f_T/10$, the AC current is approximately the same along the channel length, which is consistent with the QS approximation. However, for frequencies higher than $f_T/4$, e.g., $f_T$ or $2f_T$, the carriers do not have enough time to move from drain to source in a signal period, resulting in a departure of $i$ from the QS value, which is in accordance with the behavior observed in **Figure 13**a.

**Figure 13**c-d show the normalized magnitude and phase of $y_m = y_{dg} - y_{gd}$. It is observed that from the zero-order model to the first-order, the NQS model produces a significant improvement in the region of validity (where we have assumed here that the numerical solution provides the reference solution). This is because $y_m$ contains a right-half-plane zero for this zero-order model[119] in contrast to the left-half-plane pole in $y_m$ for the first-order NQS model (Equation 24). The upward-going magnitude predicted by the QS model at HF is unrealistic, although the phase of $y_m$ is predicted better by the QS model than the first-order NQS model; therefore, a higher order correction would be needed if the phase of $y_m$ is crucial for the targeted range of frequency according to the intended application. This discussion as well as results shown in **Figure 13**c-d are in qualitative agreement with the NQS studies conducted for conventional silicon-based MOSFETs.[128,159]

The frequency dependence of the current and power gains predicted by the different models, namely $|h_{21}|^2$, $U$, and MSG/MAG, are shown in **Figure 13**e. Differences between the QS and NQS model predictions are not significant below $f_T$, but the QS model overpredicts the gain for frequencies higher than $f_T$, highlighting the importance of including the NQS effects.

*4.2.2. Dynamic Self-Heating Model*

The equations we have presented so far to describe the drain current include a dependence on the bias point and the temperature. To analyze the impact of self-heating on HF performance, the dependence of the small-signal parameters on temperature must be studied. Here, we present a model valid for AC input voltages, assuming that the GFET presents a complex thermal impedance, $Z_{th}$, that depends on the angular frequency, $\omega$ ($= 2\pi f$), and on the different thermal paths through which the power, $P_{dis}$, is dissipated from the graphene channel. This heat crosses the different device thermal boundaries defined by the interfaces between the materials. $Z_{th}$ can be represented by various thermal *RC* networks characterized by the values of thermal resistances, $\mathfrak{R}_{th,i}$, and thermal frequencies, $f_{th,i}$, as follows:

$$Z_{th}(\omega) = \sum_i \frac{\mathfrak{R}_{th,i}}{1 + j\frac{\omega}{2\pi f_{th,i}}} \qquad (27)$$

The sum of all thermal resistances gives the total thermal resistance, which causes the increase of graphene temperature owing to DC self-heating, as addressed in §2.2.2

$$\mathfrak{R}_{th} = \sum_i \mathfrak{R}_{th,i} \qquad (28)$$

**Figure 14** shows the real and imaginary parts of $Z_{th}(\omega)$ for the simple case of two thermal *RC* networks in series (two poles), one at a thermal frequency of $f_{th,1} = 1$ kHz and the other at $f_{th,2} = 1$ GHz. The thermal frequencies highly depend on the specific GFET technology because they are a function of the heat dissipation paths. Although they have been arbitrarily chosen to illustrate dynamic SHE, they could be observed in the frequency range between $\sim 5 - 10$ MHz according to experimental measurements performed elsewhere.[54,146]

Following the temperature-dependent two-port network approach developed by Rinaldi[166] and assuming no leakage current through the gate oxides, the small-signal parameters are modified according to the following equations:

$$\begin{aligned} y_{gg}(\omega) &= y_{gg,T}(\omega) \\ y_{gd}(\omega) &= y_{gd,T}(\omega) \\ y_{dg}(\omega) &= \frac{y_{dg,T}(\omega) + cZ_{th}(\omega)y_{gg,T}(\omega)V_{GS}}{1 - cZ_{th}(\omega)V_{DS}} \end{aligned} \qquad (29)$$

$$y_{dd}(\omega) = \frac{y_{dd,T}(\omega) + cZ_{th}(\omega)(y_{gd,T}(\omega)V_{GS} + I_{DS})}{1 - cZ_{th}(\omega)V_{DS}}$$

where $c \equiv \partial I_{DS}/\partial T$ is a coefficient describing the rate of change of the current with respect to the temperature and $y_{ij,T}(\omega)$ are the small-signal parameters calculated by the QS approximation at a constant temperature, that is, assuming that the frequency is sufficiently high so the device temperature cannot follow the rapid oscillations of the electrical signal.

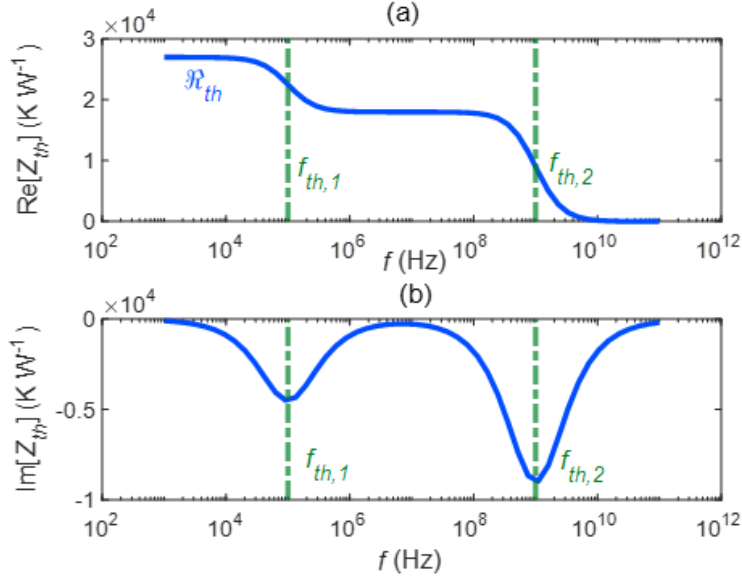

**Figure 14.** Complex thermal impedance of the graphene transistor showing two poles, each one corresponding to a heat dissipation path. **a)** Real part and **b)** imaginary part as a function of the applied frequency.

**Figure 15** shows the simulated intrinsic small-signal parameters, $y_{dg}(\omega)$ and $y_{dd}(\omega)$, for the device described in Ref.[51] at selected bias points as a function of the frequency, and assuming a two-pole thermal impedance as that presented in **Figure 14**. Above the thermal frequencies, $f_{th,1}$ and $f_{th,2}$, the thermal impedance is $Z_{th}(\omega) \approx 0$; therefore, the small-signal parameters are equal to the values at a constant temperature $y_{ij}(\omega) \approx y_{ij,T}(\omega)$. On the contrary, for very low frequencies, the graphene temperature can follow the oscillation of the electrical signal. Notably, it is very important to consider the SHE in HF performance calculation, as the QS approximation directly applied to the DC current curves assumes that the temperature varies with the bias point.

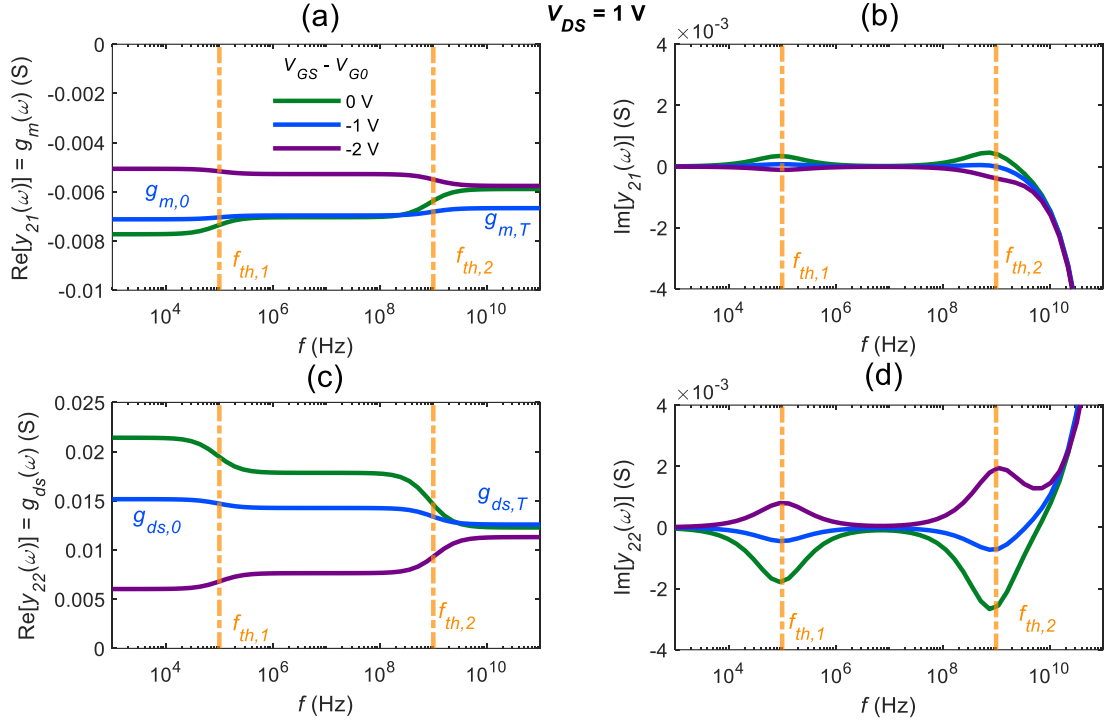

**Figure 15.** Small-signal parameters affected by SHE at the biases $V_{DS} = 1$ V, $V_{GS} - V_{G0} = 0, -1$ and, $-2$ V. **a)** Real part and **b)** imaginary part of $y_{dg}$. **c)** Real part and **d)** imaginary part of $y_{dd}$.

*4.2.3. Parasitic Network Model*

Generally, the parasitic network plays a significant role in determining the GFET performance. For on-wafer RF FETs, it is determined by the contact pads, metal interconnections, and the substrate.[167] **Figure 16**a shows the complete GFET circuit with the parasitic network connected to the intrinsic FET equivalent circuit discussed previously, as well as the contact resistances.[168]

To obtain information on internal physical mechanisms in fabricated devices for a given intrinsic model parameter to be validated for a correct description of the GFET RF performance, de-embedding techniques should be applied to remove the effect of the parasitic network. This can be done by characterizing dummy test structures developed ad hoc, which are usually fabricated on the same wafer as the device under test. The common de-embedding procedure involves measuring the S-parameters of "open," "short," and "thru" test structures and applying a mathematical procedure to subtract the effect of certain elements of the parasitic network. Therefore, the small-signal parameters of the intrinsic FET can be isolated from the de-embedded S-parameters. However, the contact resistances, $R_s$ and $R_d$, cannot be de-embedded in Schottky-like transistors such as 2DM based FETs; therefore, the parasitic resistances extracted by this method are $R_{sa}$ and $R_{da}$ (cf. **Figure 16**a). As a result, $R_s$ and $R_d$ should be extracted separately, e.g., using the transfer length method technique or an adapted Y-function based method.[169] The first one involves the fabrication and characterization of back-gated devices with different channel lengths whereas the second one is a more general approach because it can be applied to individual devices regardless the gate architecture. Once the values of $R_s$ and $R_d$ are known, their impact can be removed from the dynamic characteristics by following a matrix algebra process as shown elsewhere.[168]

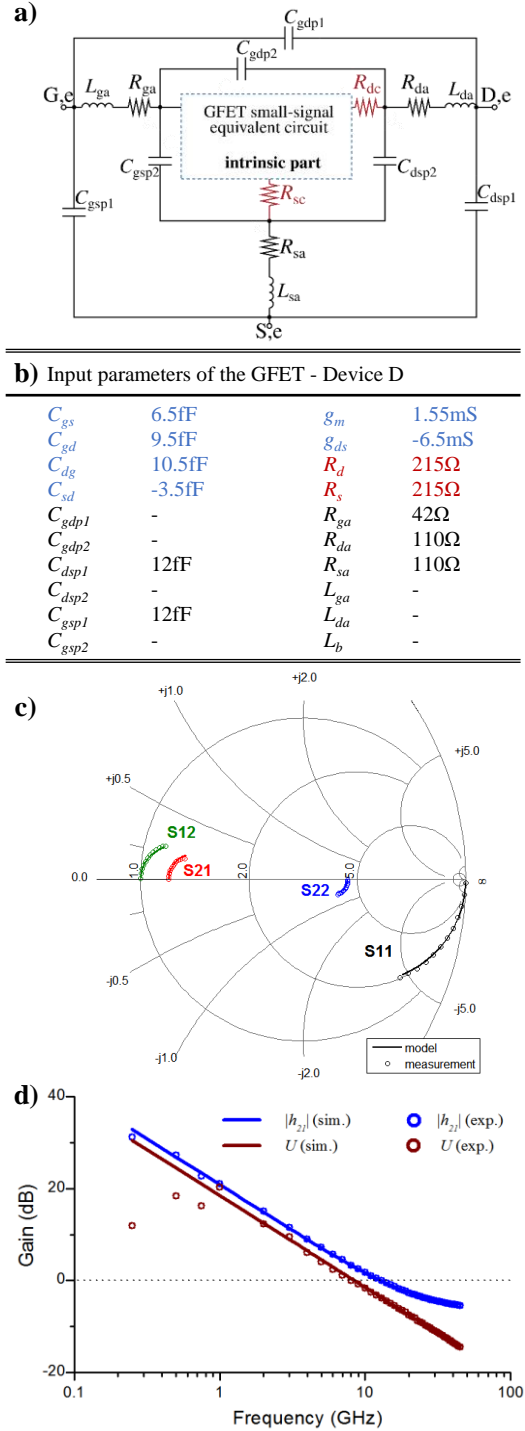

**Figure 16.** a) Typical topology of the small-signal equivalent circuit for an RF GFET. b) Parameters of the equivalent circuit for describing the experimental GFET reported in Ref.[119]. c) S-parameter measurements (circles) and simulations (lines) for the applied bias $V_{GS,e} = 0.2$ V and $V_{DS,e} = 1$ V. d) RF performance of the experimental GFET. Measured (symbols) and simulated (solid line) small-signal current gain ($|h_{21}|$) and Mason's invariant ($U$) plotted versus frequency. Reproduced with permission.[119] 2017, IEEE.

The outer parasitic capacitances (parasitic inductances) in **Figure 16**a, namely $C_{gsp1}$, $C_{gdp1}$, and $C_{dsp1}$ ($L_{ga}$, $L_{da}$, and $L_{sa}$), correspond to the electrostatic couplings (magnetic inductances) of the access layout pads. These capacitive (inductive) effects are related to the "open" ("short") test structure; hence, their values can be directly measured and/or removed from the device using the de-embedding method. In contrast, a special dedicated "open-pad-dummy" (also called "mute") structure[112,170,171] with identical layout as the active device, including the gate fingers, is required to extract and/or remove the inner parasitic capacitances, $C_{gsp2}$, $C_{gdp2}$, and $C_{dsp2}$. These parameters are associated to the electrostatic couplings between gate-to-source and gate-to-drain metal interconnect fingers (also identified as extrinsic fringing capacitances) and finger overlaps capacitance effects owing to gates wider than the channel as well as to a spurious coupling between feed and connecting lines in a finger layout. Notably, for practical purposes, e.g., the design of a functional GFET-based RF circuit, the unavoidable inner parasitics should be considered. In addition to the experimental characterization via the HF admittance parameters of the test structures,[112,170,171] the $C_{xxp2}$ values have been obtained in the literature using numerical device simulations (of the dummy structure)[172–174] as well as with compact model fitting for a correct description of measurements.[24,119,175,176] Models embracing these parasitics in GFETs are still missing in the literature. A possible approach for the correct description of effects embraced by $C_{xxp2}$ is the conformal mapping technique,[177] accounting for the fringing fields and the solution of the parallel plate capacitances within the layout as suggested in studies of other FET technologies.[161,178,179]

**Figure 16**b shows the intrinsic small-signal parameters (blue color), parasitic network parameters (black color), and contact resistances describing the experimental GFET reported elsewhere.[119] The de-embedding procedure described in Ref.[167,170,180], involving "open" test structures, has been implemented to eliminate the contribution of the parasitic network. From the de-embedded S-parameters, the intrinsic elements have been obtained through the extraction methodology proposed in Ref.[119] As a novelty, the method considers $R_s$ and $R_d$ as a part of the (augmented) intrinsic GFET model; therefore, they can be extracted on equal footing with the rest of small-signal intrinsic parameters. Measured and modeled S-parameters at $V_{GS} = 0.2$ V and $V_{DS} = 1$ V plotted together in **Figure 16**c are in good

agreement. In addition, **Figure 16**d shows the experimental extrinsic current gain ($|h_{21}|$) and extrinsic Mason's invariant ($U$), both obtained from the S-parameters measurements shown in **Figure 16**c, compared to the simulated ones obtained from the small-signal model (**Figure 16**a-b).

## 5. Noise Analysis

Graphene-based FETs have proven to be excellent contestants for forthcoming RF applications owing to graphene's exceptional characteristics that can ensure high speed performance.[181] In such circuits, noise can either be in the form of low-frequency noise (LFN) below the corner frequency $f_c$ or high-frequency noise (HFN) above the aforementioned frequency; thus, it is a crucial figure of merit.[161] LFN can deteriorate the performance of RF circuits by being up-converted to phase noise[182–185] and it can downgrade the sensitivity of chemical-biological sensors[186,187] or optoelectronic devices.[188] Conversely, HFN can be very critical for the sensitivity of an RF system and the signal-to-noise ratio at the RF regime.[189] Thus, both low- and high- frequency noise should be investigated thoroughly and modeled correctly.

There are three major mechanisms responsible for the generation of LFN in transistors: (i) carrier number fluctuation effect ($\Delta N$),[190–193] (ii) mobility fluctuation effect ($\Delta \mu$),[194] and (iii) contact resistance contribution ($\Delta R$).[161] $\Delta N$ is formed by trapping/detrapping mechanism near the oxide interface. Each individual carrier that gets captured and then released by an active trap within a few $k_B T$ from the Fermi level creates random telegraph noise (RTN),[191] which demonstrates Lorentzian shape in the frequency domain. Although experimental RTN has not been observed in GFETs, which could be because of the large area of the devices under test, the superposition of the Lorentzian spectra results in 1/f noise in longer gated channels, where the number of slow near-interfacial and border traps is adequate, if the distribution of their time constants is uniform on the logarithmic axis.[190,191] Moreover, a thermally activated process is assumed for the trapping/detrapping mechanism for $\Delta N$ 1/f noise, where time constants follow a non-radiative multiphonon model.[190,191] Fluctuations of the carrier mobility are responsible for the generation of $\Delta \mu$ model, which is described by the empirical Hooge expression, and contact resistance can also contribute to LFN ($\Delta R$).[161] Regarding CMOS processes, a number of analytical models are available in literature, which mostly refer to the $\Delta N$ effect and consider a uniform channel. They are valid only in the linear region of operation and result in a squared transconductance-to-current ratio ($g_m/I_{DS}$)$^2$ trend of normalized output noise ($S_{ID}/I_{DS}^2$). Several recent works have indicated the same mechanisms ($\Delta N$, $\Delta \mu$, $\Delta R$) responsible for the genesis of 1/f noise in GFETs.[64,195–203] Moreover, $\Delta N$ prevails as the number of graphene layers reduce whereas $\Delta \mu$ is dominant in multilayer devices.[198] The same studies depict an M-shape dependence of normalized 1/f noise data vs. effective gate voltage ($V_{GEFF} = V_{G(B)S} - V_{G(B)0}$). Although most of the 1/f noise models in GFETs are derived based on the simplified approach of uniform channel mentioned before,[196–201] we recently proposed a complete physics-based compact model,[64,202] which accounts for all the non-homogeneities as well as velocity saturation effect, that can affect 1/f noise in GFETs (§5.1). Additionally, a complete 1/f noise parameters extraction procedure has been introduced.[64,203] The model gives accurate results capturing both the M-shape dependence and the reduction of 1/f noise at high electric fields owing to the velocity saturation mechanism[64,203] when validated with data from fabricated devices.[63] The aforementioned experiments present a slightly ideal 1/f behavior (cf. Fig. 4 in Ref.[64]).

Moreover, the study of variability issues in currently immature GFET technologies is very critical for the transition to large-scale wafer fabrication and eventually to massive applications production. 1/f noise statistical deviation is a significant variable and it should be investigated thoroughly. Although low-frequency noise variance has been adequately researched in CMOS devices following empirical,[204] simplified,[205] or complete approaches,[206] there is a lack of studies regarding GFETs. Recently, we designed a complete physics-based compact model[207] describing the bias dependence of the 1/f noise variance in GFETs accurately when compared to experimental data (§5.1).[208]

For high-frequency noise modeling in GFETs, very few studies have been reported. They are mainly based on simple long-channel approaches, straightforwardly adapted from CMOS, and neither do they focus on the bias dependence of noise, nor do they consider the degenerate nature of graphene.[175,209–211] Thus, we developed a novel physics-based compact model that accurately describes high-frequency noise drain current spectral density ($S_{ID}$) for various operating conditions including the derivations in the velocity saturation effect and graphene's degeneracy (§5.2).[212] Two short channel GFETs have been measured[63] in high-frequency range to extract intrinsic $S_{ID}$ after appropriate de-embedding as well as $R_g$ and $R_c$ elimination procedures. The experimental data precisely validate the proposed model for a wide $V_{GS}$ span without using any fitting parameters and $S_{ID}$ increases towards higher-carrier densities until it saturates, similarly to MOSFETs.[213] The model is also derived for the non-degenerate case, where a significant overestimation of measurements is revealed, indicating that the degenerate nature of graphene significantly reduces high-frequency noise. In addition, noise excess factor γ, a crucial figure of merit of noise performance in RF circuits,[214] is first examined in GFETs and presented in Ref.[212].

In general, the noise models presented in §5 are extracted based on a methodology of dividing the device channel in very small slices where each one of them is a local noise source; all these local noise sources are considered uncorrelated and thus the integration of the local power spectral densities (PSDs) from source to drain results in the total channel noise.[64,189,202,203,207] Thereafter, the final compact expressions are obtained considering the chemical potential-based *I-V* model proposed in Equation 1 to 11 of §2.1. All the models are implemented in Verilog-A; thus, they can be easily integrated in circuit simulators.

5.1. Low-Frequency Noise Model

The proposed 1/f noise model's compact expressions are presented in Equations A1-A5 in Supporting Information.[64,202] The precise performance of the model is verified in Figure 17a, where normalized 1/f noise $S_{ID}f/I_{DS}^2$ is shown vs. $V_{GEFF}$ for a $W/L = 12$ μm/$100$ nm short-channel GFET with round markers representing the measurements and lines the different contributors of the model. More specifically, $\Delta N$ and $\Delta \mu$ effects are illustrated with solid lines and they are calculated by the subtraction of long-channel terms, $\Delta NA$ and $\Delta \mu A$ (dashed lines),[202] with the velocity saturation related terms, $\Delta NB$ and $\Delta \mu B$ (dotted lines), respectively;[64] $\Delta N$ terms are denoted in red, $\Delta \mu$ are denoted in blue, and $\Delta R$ are denoted in green. Evidently, $\Delta N$ is responsible for the M-shape whereas $\Delta \mu$ presents a Λ-shape and contributes near CNP. In the left subplot, the low $V_{DS} = 60$ mV regime is depicted and it is clear that the long-channel terms are dominant there, whereas the velocity saturation related ones are negligible owing to the low electric field value. On the contrary, the higher $V_{DS} = 0.3$ V region shown in the right subplot confirms that as the electric field is heightened, velocity saturation effect becomes significant; thus, $\Delta N$ and $\Delta \mu$ mechanisms and consequently total 1/f noise are reduced. The experimental data confirm the above theoretical findings, which are of outstanding significance because the unavoidable usage of high-speed short-channel devices in graphene RF circuits makes it vital to model their low-frequency noise.

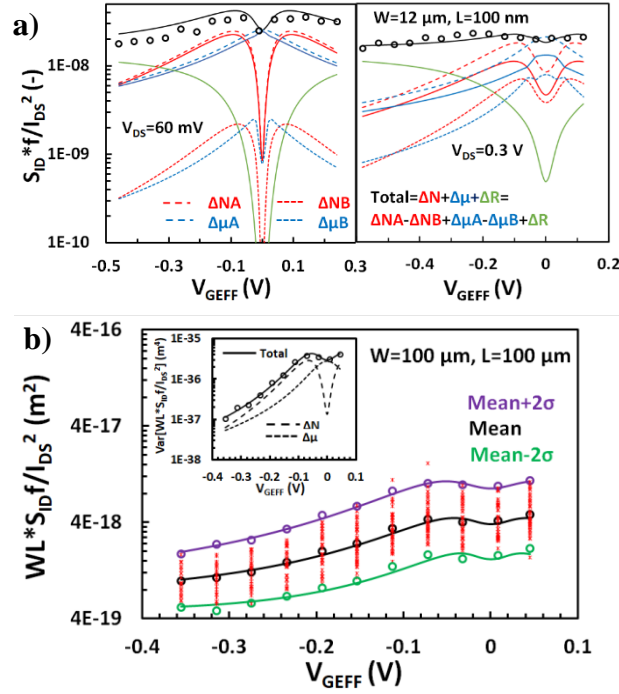

**Figure 17. a)** Drain current noise divided by squared drain current $S_{ID}f/I_{DS}^2$, referred to 1 Hz, vs. gate voltage overdrive $V_{GEFF} = V_G - V_{G0}$ at low $V_{DS} = 60$ mV in left subplot and high $V_{DS} = 0.3$ V in right subplot for a $W/L = 12\,\mu\text{m}/100$ nm GFET. Data are shown with markers whereas different noise contributors are depicted with solid, dashed, and dotted lines as well as with different colours. **b)** Normalized 1/f noise $WLS_{ID}f/I_{DS}^2$ vs. $V_{GEFF}$, for a 100 μm/100 μm GFET at $V_{DS} = 50$ mV. Measured noise from all available samples: star markers, measured ln-mean noise and its ±2-sigma deviation: open circle markers, mean and ±2-sigma deviation model: lines. (mean data and model: black, +2-sigma deviation data and model: purple, -2-sigma deviation data and model: green). Inset illustrates the variance of normalized 1/f noise $\text{Var}[WLS_{ID}f/I_{DS}^2]$ vs. $V_{GEFF}$, for the same GFET. Markers: data, solid lines: total model, dashed lines: individual contributions ($\Delta N$, $\Delta\mu$). Reproduced with permission.[64] 2019, ACS.[207] 2020, RSC.

Regarding the 1/f noise variance model, Equations A6-A10 in Supporting Information are derived and presented.[207] $WLS_{ID}f/I_{DS}^2$ 1/f noise vs. $V_{GEFF}$ for a long-channel solution-gated $W/L = 100\,\mu\text{m}/100\,\mu\text{m}$ GFET is shown in Figure 17b. Mean 1/f noise data are depicted with black markers whereas the solid lines, accounting for the mean 1/f noise model, fit these data precisely. 1/f noise data from all the samples are shown with small red markers. For the validation of 1/f noise statistical model,[207] $\pm 2\sigma$ standard deviation of normalized 1/f noise are also presented both for the model and experimental data with purple ($+2\sigma$) and green ($-2\sigma$) solid lines and markers, respectively. The proposed statistical 1/f noise model captures the dispersion of the data and its bias dependence accurately,[207] and the general picture reveals the consistency between the mean value and variance 1/f noise models. The inset of Figure 17b shows the variance of normalized 1/f noise $\text{Var}[WLS_{ID}f/I_{DS}^2]$ vs. $V_{GEFF}$ for the same GFET, where the agreement of the total model (solid lines) vs. data (markers) is consistent. Notably, the variance contributions, $\varDelta N$ and $\varDelta \mu$, shown with dashed and dotted lines, respectively, behave similarly as in the mean value 1/f noise case. Therefore, the $\varDelta N$ model provides an M-shape to 1/f noise variance as it did for its mean value, whereas the $\varDelta \mu$ model follows a $\Lambda$-shape, contributing to 1/f noise variance mainly at CNP, as it was the case for mean value 1/f noise.[207] Another important observation is that 1/f noise statistical dispersion in GFETs is not related with I-V quantities but it is caused by the deviations of the physical parameters of $\varDelta N$ and $\varDelta \mu$ mechanisms, which are the number of traps, $n_{tr}$, and the Hooge parameter, $\alpha_H$, respectively.[207]

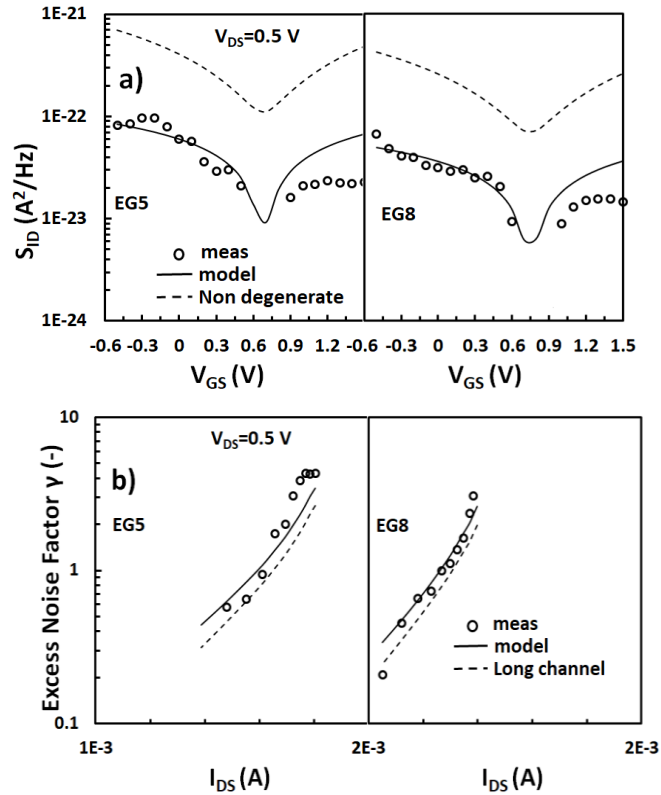

Figure 18. a) Channel thermal noise, $S_{ID}$, at 1 GHz vs. $V_{GS}$ and b) noise excess factor $\gamma$ vs. $I_{DS}$ for short channel GFETs with $W = 24$ μm and $L = 200$ nm (EG5-left subplots), $L = 300$ nm (EG8-right subplots), respectively at $V_{DS} = 0.5$ V. markers: measured, solid lines: model, dashed lines: non-degenerate model in a) and long channel model in b). Reproduced with permission[212] 2021, IEEE.

5.2. High-Frequency Noise Model

The high-frequency noise $S_{ID}$, model is accurately defined in Equations A11-A14 in Supporting Information.[212] Contrary to the LFN, where velocity saturation effect has been shown to decrease noise under high electric field conditions (cf. **Figure 17**a), it has an opposite additive effect on high-frequency noise.[209] High-frequency noise measurements have been conducted at $f = 1$ GHz, which forces the GFETs under test to operate at the quasi-static regime as their extrinsic $f_T$ is higher than $f = 1$ GHz.[63] Therefore, $S_{ID}$ is frequency-independent under such operating conditions and the attention is concentrated on its bias-dependence. **Figure 18**a presents intrinsic measured $S_{ID}$ vs. $V_{GS}$ at $V_{DS} = 0.5$ V for a $W/L =$ 24 µm/200 nm (EG5) GFET in the left and a $W/L = 24$ µm/300 nm (EG8) one in the right subplot, respectively, with markers, whereas the complete model is depicted with solid lines. $V_{GS}$ extends from high *p*- to high *n*-type carrier densities' area but the maximum $g_m$, recorded in p-type regime,[211,212] leads the attention to this specific region. Owing to the direct relation of $g_m$ with $S_{ID}$,[212] the latter is also higher there; thus, the model, which presents a symmetric behavior, accurately follows the experiments in p-type regime. Non-degenerate models are also demonstrated with dashed lines for both devices and they overestimate the experiments almost one order of magnitude, indicating that using high-frequency noise CMOS models in GFETs is not accurate. As mentioned earlier, excess noise factor, $\gamma$, is first examined thoroughly in GFETs in Ref.[212]. **Figure 18**b illustrated both experimental (markers) and simulated (lines) $\gamma$ for EG5 in the left and EG8 in the right subplot, respectively, at $V_{DS} = 0.5$ V vs. $I_{DS}$, which is now confined in the p-type region where the specific study is focused, as detailed earlier. The validation of the model with the measurements is reliable. A long-channel case is also shown with dashed lines after ignoring the velocity saturation effect, which results in a quite significant underestimation of $\gamma$. In addition to $S_{ID}$, which dominates above corner frequency $f_c$, the potential fluctuations within the channel are coupled with the gate through gate oxide capacitance resulting in induced gate noise, $S_{IG}$, and its correlation with channel noise $S_{IGID}$, both important at frequencies close or above $f_T$ at non-quasi-static regime.[161,189] These two terms increase with frequency because $S_{IG}$ is proportional to $f^2$ whereas $S_{IGID}$ is proportional to $f$. Although there are some simple long-channel approaches to model the above two contributions, they are not valid in the non-quasi-static region of operation. Currently, our model predicts a frequency independent $S_{ID}$ (white noise) that is valid for frequencies sufficiently below $f_T$, whereas $S_{IG}$ and $S_{IGID}$ have a very small effect there. The characterization and compact modeling of $S_{IG}$, $S_{IGID}$ are of critical importance for high-frequency noise at higher frequencies; this task is an ongoing research.

## 6. Conclusions

In this study, we have provided an updated report on the progress made towards the development of a modular compact modeling technology allowing DC, transient, AC, and noise analysis of arbitrary GFET-based circuits. The models have a strong physical basis and consider some non-idealities that have proven to impact in static and/or dynamic operation. The latter include extrinsic-, short-channel-, trapping/detrapping-, self-heating-, and non-quasi static-effects. The models have been validated against experimental results for the relevant operating conditions up to frequencies of some tens of GHz. We estimate that the presented modeling technology's readiness level (TRL) is 4 (validation in laboratory environment). To push the technology towards higher TRLs, more efforts are needed in different directions. First, there are some relevant physics discussed along the manuscript that deserve further modeling and/or experimental validation, such as: (i) trapping/detrapping mechanisms under dynamic operation, (ii) self-heating effects at frequencies below and close to the thermal frequency, (iii) non-quasi-static effects at frequencies near and beyond the cut-off frequency, and (iv) high-frequency noise including the gate induced noise and its correlation with channel noise at frequencies close or beyond the cut-off frequency in the non-quasi static regime. Second, moving to higher TRLs requires the successful application of state-of-the-art modeling technology to integrated circuits working in a relevant/operational environment towards the improvement and strengthening of the GFET technology by a constant feedback between fabrication technology groups, modeling groups, and circuit designers. That requires a more systematic characterization of RF building blocks such as frequency multipliers, mixers, and low-noise amplifiers, to mention a few. This would allow the full demonstration of the consistency between simulation and experiment for the relevant operation conditions. Finally, statistical variability in device characteristics is an important aspect that should be addressed in future investigations, which is critical for the transition to the large-scale wafer fabrication and eventually to massive applications production.

**Availability of compact modeling technology**

The compact modeling technology reported in this manuscript is available from the corresponding author upon reasonable request. The compact models have been protected under the Benelux Office for Intellectual Property (BOIP) with i-DEPOT number: 118389.

**Supporting Information**

Supporting Information is available from the Wiley Online Library or from the author.

**Acknowledgements**

This work has received funding from the European Union's Horizon 2020 Research and Innovation Programme under Grant Agreement No. GrapheneCore3 881603. It has also received partial funding from the Spanish Government under the project RTI2018-097876-B-C21 (MCIU/AEI/FEDER, UE); partial funding from the European Union Regional

Development Fund within the framework of the ERDF Operational Program of Catalonia 2014-2020 with the support of the Department de Recerca i Universitat, with a grant of 50% of total cost eligible. GraphCAT project reference: 001-P-001702.

**Conflict of Interest**

The authors declare no conflict of interest.

Received: ((will be filled in by the editorial staff))
Revised: ((will be filled in by the editorial staff))
Published online: ((will be filled in by the editorial staff))

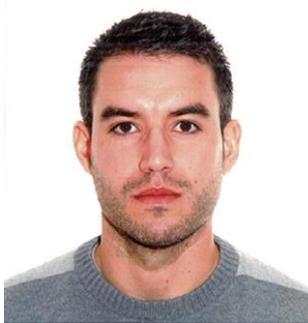

**F. Pasadas** received the Ph.D. degree in electronic engineering from Universitat Autònoma de Barcelona (UAB), Bellaterra, Spain, in 2017. From 2017 to 2021, he was a Postodoctoral Research within the Departament d'Enginyeria Electrònica at UAB, where he carried out the development of physics-based models of devices based on graphene and related materials. He is currently with the Departamento de Electrónica y Tecnología de Computadores from Universidad de Granada (Spain). His current research interests include the modeling of flexible 2D devices and the design of novel radio-frequency applications based on emergent 2D technologies.

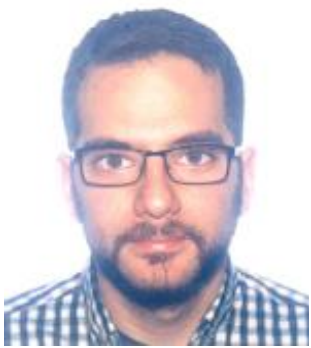

**Pedro C. Feijoo** received the B.Sc. degrees in Physics and Electronic Engineering and the M.Sc. degree in Applied Physics from Universidad Complutense de Madrid (UCM), Spain. He obtained the Ph.D. degree in Physics from UCM in 2013, in the field of deposition of high-permittivity dielectrics for silicon transistors. In 2014, he joined Universitat Autònoma de Barcelona (UAB), where he works on the simulation of graphene transistors. He was visiting researcher at Imec, Belgium, in 2010, where he worked in transistor reliability and at Aalto Yliopisto, Espoo, Finland, in 2017, where he worked in fabrication of 2D-semiconductor transistors.

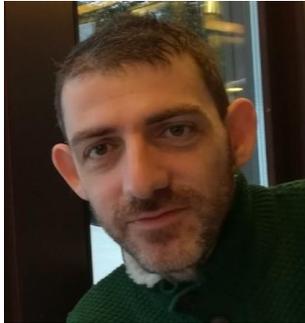

**Nikolaos Mavredakis** obtained the Ph.D. on CMOS noise characterization and compact modeling from the School of Electronic & Computer Engineering, Technical University of Crete, Chania, Greece in 2016. He had been an invited researcher with AMS, Austria (2010) and with AdMOS, Germany (2012). He had been a member of the EKV3-MOSFET compact-model development team and he was in charge for EKV3 model extraction at many projects with industry. He is currently a postdoctoral researcher at Autonomous University of Barcelona and the focus of his research is the compact modeling of noise in Graphene and 2D-material devices on the Graphene-Flagship project.

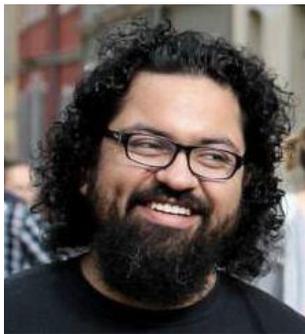

**Aníbal Pacheco-Sánchez** received the Dr.-Ing degree in Electrical and Computer Engineering from the Technische Universität Dresden, Germany in 2019 and the M. Sc. degree in Telecommunications Engineering and B. Eng. degree in Electronics and Telecommunications from the National Polytechnic Institute, Mexico, in 2011 and 2008, respectively. Since April 2019, he works as a postdoctoral researcher at Universitat Autònoma de Barcelona (Spain). His research activities embrace the characterization and modeling of emerging transistor technologies at static and dynamic operation.

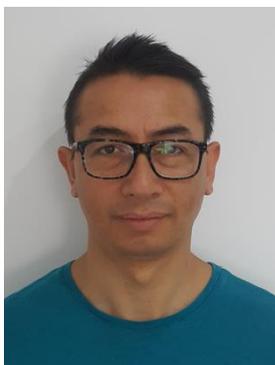

**Ferney A. Chaves** received the B.Sc. degree in Physics from the Universidad Nacional de Colombia, the Master Sc. degree from the Universidad de los Andes (Colombia) and the Ph.D in Electronic Engineering from Universitat Autònoma de Barcelona (UAB) in 1998, 2005 and 2012, respectively. From 2013 he has been working as a postdoctoral researcher at UAB mainly in the "Graphene Flagship Project" on modeling and simulation of Graphene-Metal contact resistance in graphene-based devices and electrical properties of 2D crystal-based devices such as barristors and pn heterojunctions.

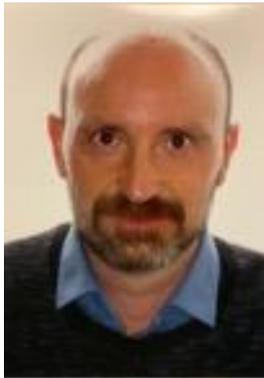 **David Jiménez** received the Ph.D. degree in electronics engineering from the Universitat Autònoma de Barcelona in 2000. He has been an Associate Professor with the Departament d'Enginyeria Electrònica, Universitat Autònoma de Barcelona, since 2004, and appointed Full Professor in 2020. His research activity is focused on compact modeling of nanoscale transistors. From October 2013 he is leading the compact modeling activities of graphene based electronic devices within the Graphene Flagship project (https://graphene-flagship.eu), which gathers over 160 academic and industrial partners from 23 countries, all exploring different aspects of graphene and related materials.

# Supplementary Information: Compact modeling technology for the simulation of integrated circuits based on graphene field-effect transistors

*Francisco Pasadas, Pedro C. Feijoo, Nikolaos Mavredakis, Aníbal Pacheco-Sanchez, Ferney A. Chaves, and David Jiménez\**

F. Pasadas, P. C. Feijoo, N. Mavredakis, A. Pacheco-Sanchez, F. A. Chaves, D. Jiménez\*
Departament d'Enginyeria Electrònica, Escola d'Enginyeria, Universitat Autònoma de Barcelona, 08193 Bellaterra, Spain
E-mail: David.jimenez@uab.cat

F. Pasadas
Departamento de Electrónica y Tecnología de Computadores, Universidad de Granada, 18071 Granada, Spain.

**S1.- 1/f Noise model:**

$$\frac{S_{ID}}{I_{Ds}^2} = \Delta N + \Delta\mu + \Delta R = \Delta NA - \Delta NB + \Delta\mu A - \Delta\mu B + \Delta R \tag{A1}$$

$$\Delta NA = \frac{4q^2(k_BT)^2N_{it}}{2kC\Delta E_\alpha WL_{eff}(\alpha k+C^2)g_{vc}}\left\{-2C\sqrt{\alpha k}\,\mathrm{atan}\left[\sqrt{\frac{k}{\alpha}}V_c\right] \pm \alpha k\ln[\alpha+kV_c^2] \pm 2C^2\ln[C \pm kV_c]\right\}_{V_{cd}}^{V_{cs}}$$

(A2)

$$\Delta NB = \frac{4q^2(k_BT)^2N_{it}\mu_{LF}k}{2C\Delta E_\alpha WL_{eff}^2(\alpha k+C^2)^3 v_{sat,0}}\left\{\frac{2C^3(C^2+\alpha k)}{C\pm kV_c} + \frac{\alpha(C^2+\alpha k)(C^2-\alpha k \mp 2CkV_c)}{\alpha+kV_c^2}\right.$$

$$\left.\mp 2C\sqrt{\alpha k}(-3C^2+\alpha k)\mathrm{atan}\left[\sqrt{\frac{k}{\alpha}}V_c\right] - (C^4-3\alpha kC^2)\ln\left[\frac{(C\pm kV_c)^2}{\alpha+kV_c^2}\right]\right\}_{V_{cd}}^{V_{cs}}$$

(A3)

$$\Delta\mu A = \frac{2qa_H}{kCWL_{eff}g_{vc}}\left\{CV_c \pm \frac{kV_c^2}{2}\right\}_{V_{cd}}^{V_{cs}} \quad \Delta\mu B = \frac{2qa_H\mu_{LF}}{kCWL_{eff}^2 v_{sat,0}}\left\{\pm\frac{k}{2}\ln[\alpha+kV_c^2]\right\}_{V_{cd}}^{V_{cs}} \tag{A4}$$

$$\Delta R = \frac{g_{ms}^2+g_{md}^2}{\left(1+\frac{R_c}{2}(g_{ms}+g_{md})\right)^2}S_{\Delta R^2}, \qquad g_{ms,d} = \frac{\mu_{LF}Wk}{2L_{eff}}\frac{C_{t,b}}{C}V_{cs,d}^2 \tag{A5}$$

where *k* is a constant (cf. Equation 5 of the main manuscript); $N_{it}k_BT/\Delta Ea$ is the slow interface trap density per unit energy [*eV⁻¹cm⁻²*] used as a 1/f noise temperature-dependent model parameter where *ΔEa* is the amplitude of the activation dispersion, $g_{vc}$ is a normalized drain current coefficient, $C=C_t+C_b$ is the sum of top and back oxide capacitances; $\alpha$ is a residual charge related term; $\alpha_H$ is the unitless Hooge 1/f noise model parameter, $S_{\Delta R^2}$ is the *ΔR* 1/f noise model parameter [$\Omega^2/Hz$]; and $g_{ms,d}$ are the source and drain transconductances, respectively [1], [2].

**S2.- 1/f Noise Variance model:**

$$\text{Var}\left[WL\frac{S_{ID}}{I_{DS}^2}\right] = \text{Var}[\Delta N] + \text{Var}[\Delta\mu] = \text{Var}[\Delta NA] + \text{Var}[\Delta NB] + \text{Var}[\Delta\mu A] + \text{Var}[\Delta\mu B] \quad (A6)$$

$$\text{Var}[\Delta NA] = \frac{16q^4L^2(k_BT)^2N_{it}N_{tcoeff}}{\Delta E_\alpha CWL_{eff}^3 g_{vc}} \frac{\sqrt{k}}{8(C^2+\alpha k)^5} \left\{ \mp \frac{4C^4\sqrt{k}(C^2+\alpha k)^2}{(C+kV_c^2)^2} \mp \frac{16C^3\sqrt{k}(C^2-2\alpha k)(C^2+\alpha k)}{C \pm kV_c} \right.$$

$$\mp 2\alpha\sqrt{k}(C^2+\alpha k)^2 \frac{(\alpha^2k \mp C^3V_c - 3\alpha C(C \mp kV_c))}{(\alpha+kV_c^2)^2}$$

$$-\frac{\sqrt{k}(C^2+\alpha k)(5C^2V_c \pm 2\alpha C^3(12C \mp 17kV_c) + 3\alpha C^2k(\mp 8C + 3kV_c))}{\alpha + kV_c^2}$$

$$+\frac{3C}{\sqrt{\alpha}}(C^6 - 25\alpha C^4k + 35\alpha^2C^2k^2 - 3\alpha^3k^3)\operatorname{atan}\left[\sqrt{\frac{k}{\alpha}}V_c\right]$$

$$\pm 24\sqrt{k}(C^6 - 5\alpha C^4k + 2\alpha^2C^2k^2)\ln[C \pm kV_c]$$

$$\left. \mp 12\sqrt{k}(C^6 - 5aC^4k + 2\alpha^2C^2k^2)\ln[\alpha + kV_c^2] \right\}_{V_{cd}}^{V_{cs}}$$

(A7)

$$\text{Var}[\Delta NB]$$

$$= \frac{16q^4L^2(k_BT)^2N_{it}\mu_{LF}N_{tcoeff}}{\Delta E_\alpha CWL_{eff}^4} \frac{k^2}{12(\alpha k+C^2)^7 v_{sat,0}} \left\{ \frac{4C^5k(C^2+\alpha k)^3}{(C \pm kV_c)^3} + \frac{6k(C^2+\alpha k)^2(3C^6-5\alpha C^4k)}{(C \pm kV_c)^2} \right.$$

$$+\frac{24C^3k(C^2+\alpha k)(3C^4-12\alpha kC^2+5\alpha^2k^2)}{C \pm kV_c} - \frac{2\alpha^2(C^2+\alpha k)^3(C^4-6\alpha kC^2+\alpha^2k^2 \mp 4C^3kV_c \pm 4\alpha k^2CV_c)}{(\alpha+kV_c^2)^3}$$

$$+\frac{2\alpha C(C^2+\alpha k)^2(3C^5-30\alpha kC^3+15C\alpha^2k^2 \mp 13C^4kV_c \pm 30\alpha k^2C^2V_c \mp 5V_c)}{(a+kV_c^2)^2}$$

$$-\frac{3C(C^2+\alpha k)(2C^7-48\alpha kC^5+90\alpha^2kC^3-20\alpha^3k^3C \mp 11C^6kV_c \pm 79\alpha k^2C^4V_c \mp 65\alpha^2k^3C^2V_c \pm 5\alpha^3k^4V_c)}{\alpha+kV_c^2}$$

$$\mp \frac{15C\sqrt{k}}{\sqrt{\alpha}}(C^8 - 28\alpha kC^6 + 70\alpha^2kC^4 - 28\alpha^3kC^2 + \alpha^4k^4)\operatorname{atan}\left[\sqrt{\frac{k}{\alpha}}V_c\right]$$

$$\left. + (-C^8 + 7\alpha kC^6 - 7\alpha^2kC^4 + \alpha^3k^3C^2)\ln\left[\frac{(C \pm kV_c)^2}{\alpha+kV_c^2}\right] \right\}_{V_{cd}}^{V_{cs}}$$

(A8)

$$\text{Var}[\Delta\mu A] = \frac{4q^2L^2\alpha_H}{k^2CWN_{aH}L_{eff}^3 g_{vc}} \left\{ \frac{C\operatorname{atan}\left[\sqrt{\frac{k}{\alpha}}V_c\right]}{\sqrt{\alpha k}} + 0.5 \pm (\alpha + kV_c^2) \right\}_{V_{cd}}^{V_{cs}} \quad (A9)$$

$$\text{Var}[\Delta\mu B] = \frac{4q^2L^2\alpha_H\mu_{LF}}{k^2CWN_{aH}L_{eff}^4 v_{sat,0}} \left\{ \mp \frac{0.5k}{\alpha+kV_c^2} \right\}_{V_{cd}}^{V_{cs}} \quad (A10)$$

where $N_{tcoeff}$ and $N_{\alpha H}$ are the 1/f noise variance model parameters [3].

**S3.- Channel Thermal Noise model:**

$$S_{ID} = S_{IDA1} + S_{IDA2} + S_{IDC} \tag{A11}$$

$$S_{IDA1} = 4K_B T U_T k \mu_{LF} \frac{W}{C g_{vc} L_{eff}} \left[ \pm \frac{\alpha C V_c^2}{2k} + \frac{\alpha V_c^3}{3} \pm \frac{C V_c^4}{4} + \frac{k V_c^5}{5} \right]_{V_{cd}}^{V_{cs}} \tag{A12}$$

$$S_{IDA2} = 4K_B T U_T k \mu_{LF}^2 \frac{W}{C v_{sat,o} L_{eff}^2} \left| \left[ \frac{k V_c^3}{3} \right]_{V_{cd}}^{V_{cs}} \right| \tag{A13}$$

$$S_{IDC} = 4K_B T U_T k \mu_{LF}^3 \frac{W}{v_{sat,o}^2 L C^2 L_{eff}^2} (V_{cs} - V_{cd}) \left[ \pm \frac{k^2 V_c^4}{4} \right]_{V_{cd}}^{V_{cs}} \tag{A14}$$

where $S_{IDA1}$ is a long channel term and $S_{IDA2}$, $S_{IDC}$ are the velocity saturation induced terms [4].